\documentclass[10pt]{article}
\usepackage{graphicx}

\usepackage{epsfig}
\usepackage{latexsym}
\usepackage{amsmath}

\begin{document}

\title{\Large \bf  Zipf-Mandelbrot-Pareto model   \\for  co-authorship popularity }

\author{ Marcel Ausloos$^{1,2,}$\footnote{previously at GRAPES, ULG, Li\`ege, Belgium; email: marcel.ausloos@ulg.ac.be   } 
   \\   \\
$^1$ eHumanities group,
Royal Netherlands Academy of Arts and Sciences, \\Joan Muyskenweg 25, 1096 CJ Amsterdam, The Netherlands \\ \\ 
$^2$ R\' es. Beauvallon, rue de la Belle Jardini\`ere, 483/0021\\
B-4031, Li\`ege Angleur, Euroland }

 \date{\today}
\maketitle
 \vskip 0.5truecm

\begin{abstract}  
 
Each co-author  (CA) of any scientist can be given a rank ($r$) of importance according to the number  ($J$) of joint publications which the authors have together.  In this paper, the Zipf-Mandelbrot-Pareto law,   i.e.  $ J \propto 1/(\nu+r)^{\zeta}$  is shown to reproduce the empirical relationship between   $J$ and $r$ and  shown to be preferable to a mere power law, $ J \propto 1/r^{\alpha} $.  
The CA core value, i.e. the core number of CAs, is unaffected, of course.  The demonstration is made on data for  two authors, with a high number of joint publications,   recently  considered by Bougrine (2014) and for 7 authors, distinguishing between their "journal" and "proceedings" publications as suggested by Miskiewicz (2013).
 The  rank-size  statistics is discussed and  the $\alpha$ and $\zeta$ exponents are compared.   The correlation coefficient is much improved ($\sim$ 0.99, instead of 0.92).  There are marked deviations of  such a co-authorship popularity law depending on sub-fields.      
 On one hand, this suggests an interpretation of the parameter $\nu$. On the other hand, it suggests a novel model on the (likely time dependent) structural   and publishing properties of research teams. 
  
  Thus, one can   propose  a scenario for how a research team is formed and grows. This is based on a hierarchy  utility concept,    justifying the  empirical Zipf-Mandelbrot-Pareto law, assuming a simple form for the CA publication/cost ratio,
  $c_r = c_0\: log_2 (\nu+r)$. In conclusion, such a law and model   can suggest  practical applications on   measures of research teams. 
 
In   Appendices,  the    frequency-size  cumulative distribution function is discussed  for two sub-fields, with other technicalities

 \end{abstract}

{\bf Keywords} : ranking; Zipf-Mandelbrot-Pareto law; power laws; co-authorship; research topics; peer review journals; proceedings; co-author core

\maketitle

\section{Introduction  }\label{sec:intro}

One basic question in scientometrics concerns the number of relevant co-authors  (CA) for  the set of publications of   some principal investigator (PI).  This leads to examine the  CA core index  (Ausloos  2013), $m_a$,  introduced on several grounds, including a statistical framework. Recall that   the   $m_a$-index      is   deduced from a plot of the number ($J$) of     joint   publications (NJP)  of a PI  with CAs ranked according to their rank ($r$) of  importance; $r=1$ being the most prolific CA with the PI.    It has been found  (Ausloos  2013) that
\begin{equation}\label{eq1}
  J(r) = a/r^{\alpha} ,
  \end{equation} 
   which holds  best  when there  are enough  $J$ and CA. It holds as well if   the publication list is broken into  specific types of publications, i.e.   in sub-fields (Bougrine 2014) or peer-review journals and proceedings (Miskiewicz    2013). Miskiewicz   (2013) has much discussed   the exponent $\alpha$  value which turns out to be  scientist dependent, whence is a valid criterion for team work assessment. 
However, neither  Miskiewicz  nor Ausloos  have  provided a statistical physics-like model or  any argument  for the findings.

 In the present paper,  such a model is presented as an adaptation of Mandelbrot model for text appraisal.  In order to present arguments in such a favor, the  Bougrine (2014) data is  first reexamined along the 3-parameter Zipf-Mandelbrot-Pareto  (ZMP) law  (Zipf  1949, Mandelbrot 1960, Pareto 1896, West \& Deering 1995)  
 \begin{equation} \label{ZMeq3}
J(r)=b/(\nu+r)^{\zeta}.
\end{equation}
Next, several other co-authorships are examined distinguishing between their publications in (peer review) journals or proceedings and  in a few cases the research sub-fields. 

Obviously, 
the ZMP distribution is a natural generalization of an inverse power law.  
   The ZMP law  leads to a 
curvature at low $r$  in a log-log plot (if $\nu$ is positive\footnote{Necessarily,  $-1 \le \nu $, since $r \ge 1$})  and presents an asymptotic power
 law behavior  at large $r$.  
Both $\alpha$ and $ \zeta$ exponents must be greater than 1 for the distributions to be well-defined,  greater than 2 for the mean to be finite, and  greater than 3 for the variance to be finite. 
Note that the ZMP  distribution has been applied in many contexts,   e.g.,  in  studies of   scientific citations (Tsallis \& Albuquerque 2000).  
In fact, another important aspect of the ZMP distribution is that it arises  in the context of  generalized statistical physics (Tsallis  1988).  

Fitting the ZMP law   is much more troublesome than fitting the Zipf hyperbolic law (Fairthorne 1969, Haitun 1982,  Iszak 2006).  Thus, a variant of the ZMP law, i.e. the 4-parameter  relation
 \begin{equation} \label{ZMeq4}
J(r)=c/(\eta+ \lambda r)^{\mu} 
\end{equation}
has been also examined, for comparing the resulting precisions of the  respective fits. Since nothing drastic has been found in the present case, some data and analysis is left in Appendix   A  
for   completeness.

In order to argue in favor of  a ZMP model for co-authorship popularity, 
\begin{itemize} \item two PI cases are here below examined: one is  H.E. Stanley (HES): he has more than 1400 publications, is a guru of statistical mechanics; his group website distinguishes between activities in several research sub-fields. HES  lists  more than 500 CA.   The other is M. Ausloos (MRA); he has published a little bit less than 600 papers in international journals or proceedings with reviewers;  MRA  has more than 300 CA.   Both PIs have worked in different research fields, sometimes overlapping. Remarkable fits   to the hyperbolic  law, Eq.(1), are available  (Miskiewicz    2013, Bougrine 2014)  for both PIs, - at least in the central $J(r)$ region.   Deviations  mainly  exist in the extreme regions.  Thus, it seems of interest to reexamine the data in a more parametrized way, along the ZMP law;
  \item  several other authors (7) CA lists are also examined, beside  HES and MRA, - distinguishing between their publications in journals or proceedings; 
  \item  finally,   the largest publication lists, i.e. those of HES and MRA, can be broken into several  research sub-fields; the ZMP law validity and  fit parameter values  are examined in such cases.
\end{itemize} 

After       this brief introduction serving as basic arguments,    the  $J(r)$  data  of the co-authorship features is   rank-size analysed, in Sect. \ref{sec:datasetanal}, for both MRA and HES,  and for 7 other authors in Sect.\ref{proceedingspeereview}  and in Sect.\ref{sub-fields} along the above the simple empirical laws, Eqs.(\ref{eq1})-(\ref{ZMeq3}), distinguishing between the different types of publications mentioned here above.

In Sect. \ref{sec:datasetdiscussion}, some discussion on the statistical   aspects of these illustrative cases are presented, stressing the parameter $\nu$, beside the characteristic exponents, either  for co-authored papers in journals or proceedings, Sect. \ref{sec:datasetdiscussionwjp} or according to envisaged sub-fields, Sect. \ref{sec:datasetdiscussionsub-fields}.  It is argued that the ZMP  law should be further considered from a theoretical point of view., when examining    so called king, vice-roy, queen and harem effects, in Sect. \ref{sec:KVRQH}. Therefore,    a novel model  is presented on the  structural   and publishing properties of research teams, in  Sect.  \ref{sec:model}. Thus, one can   propose  a   (rather common sense, $\sim$ Ockam's razor mode) scenario for how a research team is formed  around a PI and grows with preferred CAs. This is based on a hierarchy  utility concept  assuming a simple form for the CA publication/cost ratio.

Some summary and a short conclusion are found  in Sect. \ref{sec:conclusions}. 
 
As mentioned, practical considerations on fit precision through either  Eq.(\ref{ZMeq3}) or  Eq.({\ref{ZMeq4})   are found in  Appendix A. 
  A thorough discussion on  generalizing the main text considerations to  "sub-cores" of  co-authors  is  found  in Appendix B. 
  
  {\it In fine},  another type of display  that the   rank-size plots is of interest in order  to remain in line with somewhat more conventional plots in bibliometrics, i.e.   frequency-size plots.
For   a brief  comparison with other types of displays, and for some completeness, the    cumulative distribution function (CDF) of the number of co-authors  as a  function of NJP,  is reported as examples and commented upon, for the case  of  two sub-fields of MRA, in Appendix C.  

\section{The data,   fits,  and  statistical analysis} \label{sec:datasetanal}

     \begin{figure}
  \includegraphics[height=9.8cm,width=10.8cm]{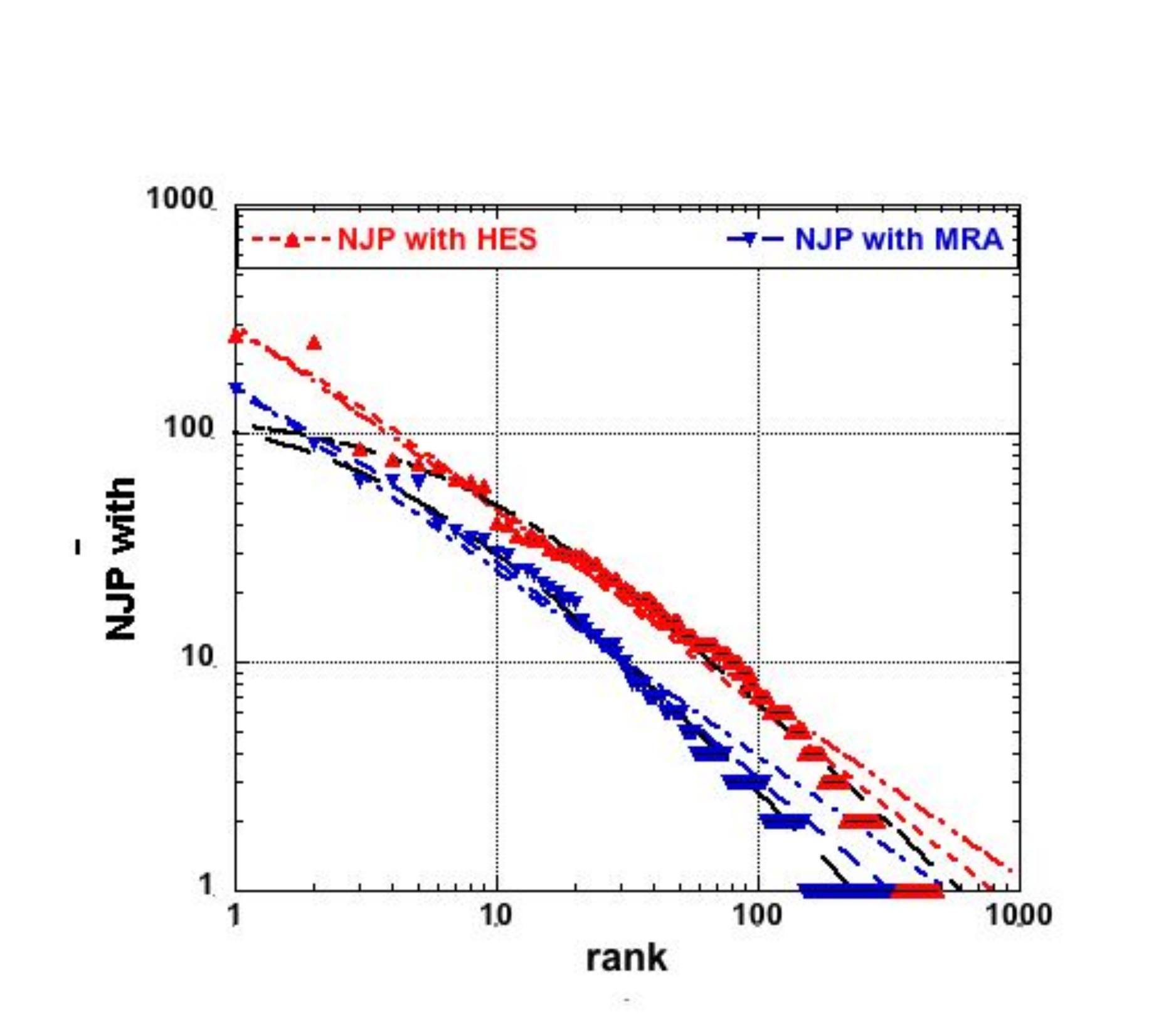}                                                                                              
\caption{  Log-log scale display  of the  rank-size relationship  between the  total number of joint publications 
 either with HES or with MRA,  for  co-authors ranked by "decreasing importance"; best fits (color lines) are shown for the power law and ZMP  law. Note that  black dash lines are fits correspond to ZMP taking into account a king and vice-roy effect, i.e.  removing  2  data points at low rank as outliers   }
\label{fig:MRAHES}
\end{figure}

The $whole$ ($w$) joint publication list  for the two prolific PIs with many CAs, i.e. HES and MRA,   has been  first studied,  - the data  being fitted with    the best power law and best ZMP fits. The parameter values are reported in Table \ref{Table3peerproceedings}. The fits are rather remarkably good, both for the power law and the ZMP laws, with an R$^2$ above 0.94. The exponents are close to 1.0, as expected. The parameter values and the regression fit coefficient R$^2$ as given for the different cases have been rounded up to significant decimals.     

 Note that it is not surprising that   all the numerical results slightly  differ from those of  Bougrine  (2014). This is likely due to the type of algorithm used in these non linear fits.  Bougrine (private communication) used a classical Least Squares curve fit, on log-log plots.  In the present study,  the fit is a non-linear one that uses the Levenberg-Marquardt algorithm.  However, there is no disagreement on the  qualitative conclusions.   In fact, as mentioned,  fitting the ZMP law   is much more troublesome than fitting the Zipf hyperbolic law (Fairthorne 1969, Haitun 1982,  Iszak 2006).
However,  the error bars    on the parameter values, calculated from several simulations with different initial conditions, are less than 10\%, - usually less 5\%.     For completeness,  note that the quality of the fits were examined through the statistics of the errors (Jarqe \& Bera 1980), i.e. whether   the error distributions were relatively small and randomly distributed on the both side of the means.
The worse cases occur when    king\footnote{The effect occurs when the data is upsurging at low rank and has been  so called when examining city size by  (Laherr\`ere \&   Sornette 1998);  it seems to have been emphasized first by Jefferson (1939), also when studying city sizes.}  and/or  queen\footnote{The effect  has been  so called when examining co-authorship sizes size by Ausloos (2013); it occurs when the data is flattening at low rank}  effects, see details below,  are superposed to a power law tail, or  when a strong exponential cut-off is present on the tail. Note also that   fits in such cases necessarily imply large error bars on the amplitudes  $a$, $b$, and $c$, since the  number of CA having a given NJP can be large itself,  the more so at high rank, when the number of publications with  the PI, usually 1 or 2,  is the same for many CAs.

It is well seen,  on Fig.\ref{fig:MRAHES},  that a plot of the (whole) NJP data  reveals a marked king effect  both for MRA and HES, -  even a king and vice-roy effect for HES. Thus, to emphasize the latter effect, the two lowest rank data points, the king  and the vice-roy, have been removed before new fit attempts. The results are shown in Table 1 also, in HES* and  MRA* lines.  Observe that  the major changes occur in the amplitudes, $\nu$ and $b$, - not in the fit exponents which rather indicate the asymptotic regime; note that $R^2$ goes from 0.92 to 0.99 for HES.

 \subsection{Peer-review journals and  proceedings }\label{proceedingspeereview} 
 
 A reviewer  of the first submitted version of this report pointed out that (I quote) "picking such singular authors allow a better statistic,  ...".  Indeed that is   a relevant point. In fact, one tends in physics to look for large numbers,  in view of finding some "universality", - but not many authors have so many papers as HES or MRA.   For analysis consistency, I had taken the data in Bougrine (2014) publication. 
 Nevertheless,  it is of interest to look for less prolific authors, as pointed out by the reviewer.  
Thus, I have requested from reliable  colleagues, having worked  in statistical and/or condensed matter physics, some pertinent data. However, as an additional  "constraint",  I have  asked such colleagues not to break their publication list into sub-fields\footnote{that would have led to too few papers per field, and it would have been nonsense to do some meaningful fit thereafter.}, as Bougrine (2024) did,  but  rather to break their publication list into "peer-review journals" and "peer-review (or not) proceedings", as Miskiewicz  (2013) has investigated.   No need to say that proceedings papers are often peer reviewed as well, but the distinction, as the one Miskiewicz proposed, is surely obvious to or acceptable   by many.

Therefore, this leads to   Table \ref{Table3peerproceedings}   and to many possible  plots, - only a few are shown such that some emphasis is placed on major features or findings,  Figs. \ref{Plot2DSJMKAPloloZMP3},   \ref {Plot2NJP6lolo}, and  \ref{Plot14PCjpjp}.  Note that
 PIs  are ranked  according to NDCA rather than NJP, for better emphasis of PI popularity hierarchy than productivity, in  Table \ref{Table3peerproceedings}.

In Fig.  \ref{Plot2DSJMKAPloloZMP3}, a  log-log scale display   of  the whole set of  joint publications (NJP$^{(w)}$)  number by various  (3) authors (see insert: DS, KK, JMK, AP)  with different production outputs and  is shown with co-authors (CA) ranked by decreasing Óimportance". Note that  $R^2 \simeq 0.98$, except for JMK (seeTable 1).       Observe also a king effect for JMK, but the data has not been further analyzed, as for HES and MRA; if done, i.e. removing the king, it should surely markedly improve the $R^2$ coefficient, actually 0.95.

In  Fig.   \ref {Plot2NJP6lolo},  co-authors are ranked on the x-axis in decreasing order of their number   of publications, in proceedings,  with the 2 two "main" PIs  and 5 other PIs  in statistical physics.  It is seen that the ZMP law holds   for such a type of publications, as well, - with marked king and queen effects.

In contrast,  the display of Fig. \ref{Plot14PCjpjp} shows a comparison of various types of publications by one author (PC), pointing out to a marked king effect, whatever the publication type, and another comparison  of publication types with co-authors (PVdB)  where marked queen effect is seen. At this point, it is sufficient to know that PVdB is an experimentalist working in condensed matter.

Thus, the various types of behaviors have been illustrated for various types of publications, indicating the ZMP framework  interest. The  values of  the fit parameters,   reported inTable   \ref{Table3peerproceedings},  can be discussed  in Sect. \ref{sec:datasetdiscussion}.

   \subsection{Sub-field effects}\label{sub-fields}
Next, let it be recalled that  the   NJP  analysis was  suggested by  Bougrine (2014) to be breakable into sub-fields:  e.g.,    into   12  appropriate sub-fields,  for HES, here called $s_i$,   and into  8  sub-fields,   in the case of MRA, here called $a_i$, - according to the web site  of such authors.  For coherence, the same  set of sub-fields has been used here. Let it be recalled that for justified statistical purposes, Bougrine has also merged some $s_{3}$    and  $s_{6}$ into some $s_{63}$       and some $a_{4}$  and  $a_{5}$ into  some $a_{54}$. The same is done here; names of sub-fields are presently irrelevant. Two  specific  cases, i.e. $a_2$ and $a_7$ (they  refer  to work on Magnetic Materials and   Superconductivity,    respectively)  will be examined in Appendix B and in  Appendix C,  as mentioned in the Introduction,  i.e. for comparing the size-rank method with the size-frequency method.

The hyperbolic law, Eq. (1)  has been studied on all $s_i$ and $a_i$ data;  numerical results reported in Table \ref{Tablestat3}.  The law is  often well obeyed,  since R$^2$ $\in \sim [0.88, 0.99]$, - except for $s_4$, in which case  R$^2\sim0.787$.     Observe that the cases with the "worse" exponent, i.e. $\alpha$ rather away from +1 in $s_9$ and $a_6$, correspond to the $oldest$  sub-fields of investigations  both for  HES, i.e. phase transitions and critical phenomena, and for  MRA, i.e., kinetic growth and spin models, - in fact, two closely related sub-fields! An open question is whether  such a variation is due to time effects or to sub-field distinctions.

 The 3 parameter ZMP (3-ZMP)\footnote{ so does  the 4 parameter ZMP (4-ZMP) law, see Appendix A} law necessarily leads to a better  R$^2$, except for   $s_3$  and $a_5$ which are almost unchanged with respect to the hyperbolic law. Recall that   they belong to  data sets with very low numbers of CAs and NJP.  All other R$^2$ values move close to 0.98 or above from $\sim$ 0.92 on average for the power law.
 
The values of  the fit parameters will be discussed  in Sect. \ref{sec:datasetdiscussion}.

  \begin{table}  \begin{center} 
 \begin{tabular}{|l|c|c|c|c| c|c|c|c|c|c| c| }
   \hline
 \multicolumn{4}{|c |}{ }&\multicolumn{4}{|c|}{ 3 param. ZMP, Eq.(\ref{ZMeq3})}&&\multicolumn{3}{|c|}{ power law, Eq.(1)}   \\ 
\hline
PI&NJP&NDCA& $<$NCA$>$  
&  $b$& $\nu$& $ \zeta$&R$^2$&&$a$&$\alpha$&R$^2$    \\
   \hline   
\hline 
HES (w)&1148&592 &6.57 &335.0 &0.10 &0.83 &0.92  &- &314.&0.81&0.92  \\
 HES$^{*}$ (w)	&*&*&*&1128.0&7.384&1.098&0.987&-&246.9 &0.765& 0.96\\
HES (j)&791&568 &4.67 &186.0&-0.10! &0.78 &0.94 &-& 200&0.80& 0.94 \\
HES (p)&296&242&5.15&255.4  &1.34&1.0 &0.97&- &119.&0.76&0.95 \\
\hline 
MRA  (w)&599&319&4.87 		&276.25 &0.88 & 0.97&0.99  	&- &161.9&	0.81 & 0.98 \\
MRA$^{*}$ (w)	&*&*&*& 723.3&4.149&1.203&0.988&-&207.2 &0.894&0.96  \\
MRA  (j)&359&273&3.86		& 183.4&1.02 &0.95 &0.99  	&-&101.7 &0.78&0.98  \\
MRA  (p)&164& 128&3.04 	&87.78&0.58&0.94&0.99		&-&59.67 &	0.81&0.98\\
\hline 
DS (w)&612&280&2.72	&953.1 &10.56& 1.38&0.98 &-&43.18&0.62&0.85  \\
DS (j)&374&268& 2.58		&637.0 &9.52 &1.31 &0.98  &-&38.74&0.62&0.86  \\
DS (p)&59&46&1.57 		&7.04&-0.01!&0.58&0.96&- &7.08&0.58&0.96\\
  \hline
  PVdB (w)&99&129&4.25&529.2&3.15&1.44&0.99&-&76.4&0.82&0.94\\
PVdB (j)&73&105 & 3.82&296.8&2.485&1.38&0.975&-&58.3&0.83&0.94  \\
PVdB (p)&26&56 & 2.14&40.7&1.47&1.07&0.98&-&16.6&0.75 & 0.96 \\
  \hline
KK (w)&172&93 &6.01&1085.1 &6.21 &1.67 &0.985  &-& 47.02&0.735&0.90  \\
KK (j)&144& 85&5.66 &1191.4&6.77 &1.72 &0.986  &- &41.79&0.732&0.90\\
KK (p)&28&28 & 2.79&10.20&0.65 &0.765 &0.946& -  &7.21&0.635&0.94\\
   \hline
  AP (w)&111& 47&2.87 &35.61&0.72&0.98&0.98&-&21.81&0.80&0.98  \\
AP (j)&79& 45&2.62 &42.21 &1.27 &1.07  &0.98 &-&18.50&0.78 &  0.96\\
AP (p)&9&11&1.55 &4.74 &0.21 &0.78 &0.93  &- &4.11 &0.71 &0.93  \\
 \hline
  JMK (w)&60&41&2.71 &27.45 &1.57 &0.87 &0.95  &-&13.17 &0.63&0.94\\
JMK (j)&28&35 &1.71 &23.50 &2.54 &1.00 &0.93 &-&7.165&0.60&0.89  \\
JMK (p)&19&25&2.04 &15.47&1.50&0.88&0.97&- &7.34 &0.61 &0.95  \\
  \hline
  PC (w)&33&32&3.09 		&8.18 &-0.90! &0.56 &0.99  	&-&28.32&1.25 & 0.94 \\
PC (j) &22&19&2.68 		& 6.59&-0.78! &0.74  &0.99 	&-&19.53&1.41&0.98 \\
PC (p)&10&24&2.0 		&3.89 &-0.91!&0.40 &0.96 		&- & 8.93&0.78&0.89\\
 \hline
JM (w)&27&15 &1.80 &2.24&-0.97! &0.49&0.99&-&11.75&1.87&0.92  \\
JM (j)&15& 13&1.77 &2.23 &-0.95! &0.515&0.99&-&9.70&1.63&0.92  \\
JM (p)&3& 3& 1.33&(-) &(-) &(-) &(-)  &- &1.96&0.75&0.93  \\
  \hline
\end{tabular}
\caption{ Summary of   fit parameters to NJP data corresponding to   Figs.\ref{Plot2DSJMKAPloloZMP3}-\ref{Plot14PCjpjp}   for various PIs,  distinguishing between their whole (w) publication list and their papers  published in   journals (j) or in  proceedings (p); the  parameter values  correspond to the various  formulae discussed in the text, Eqs.(\ref{eq1})-(\ref{ZMeq3});      the regression fit  coefficient R$^2$ is given  for the different cases; data has been rounded up to significant decimals; error bars $\le$ 10\%;
 NDCA : number of different CAs; 
 $<$NCA$>$: number of co-authors on average on 	 a paper by  the "main" author (PI);  
 for HES* and MRA*, see text}\label{Table3peerproceedings}
 \end{center}
 \end{table}

\begin{table}  \begin{center} 
 \begin{tabular}{|c|c|c|c|c| c|c|c|c|c|c| c|c|c|c|c|c|   }
   \hline
  \multicolumn{1}{|c|}{ }& \multicolumn{4}{|c|}{ 3 param. ZMP, Eq.(2)}&\multicolumn{3}{|c|}{ 2 param. law, Eq.(1)}  \\ 
\hline
 NJP &  $\nu$& $b$& $\zeta$&R$^2$&$a$&$\alpha$&R$^2$    \\
  \hline
\hline 
\hline    $s_1$	 &   0.674&27.493&0.838&0.988& 18.71&0.707&0.983\\
\hline    $s_2$	 &  7.441 &2434&2.074&0.978& 33.05&0.740&0.885\\
\hline    $s_4$	 &  12.12& 4.6 10$^5$& $q.pw.t$& 0.940& 49.935&0.841&0.787\\
\hline    $s_9$	 &   3.813&35.13&0.859&0.961& 10.801&0.522&0.930\\
\hline    $s_{11}$ &  0.956&25.692&0.989&0.974& 13.801 &0.754&0.964\\ 
\hline
\hline    $s_5$	 &  	 6.77&2003.9&$1.68$&0.979& 76.78 &0.737&0.886  \\
\hline    $s_7$	 &  0.196&74.116&1.054&0.994& 61.99 &0.982&0.994  \\
\hline    $s_8$	 &  1.172&90.31&1.128&0.982& 40.13&0.844&0.972  \\
\hline    $s_{10}$	 &   4.221&579.37&1.427&0.982& 63.09&0.756&0.920  \\
\hline    $s_{12}$	 &  	3.157&357.47&$1.19$&0.987 & 77.70&0.731&0.928  \\
\hline 
\hline
\hline    $a_ 1$	 &   2.59&28.478&1.088&0.939& 7.694&0.637&0.908  \\
\hline    $a_ 2$	 &  	20.77&2.9 10$^4$&$q.pw.t$&0.980& 22.755&0.59&0.823 \\
\hline    $a_ 3$	 &   0.365&17.029&0.893&0.971& 13.164&0.793&0.969  \\
\hline    $a_ 6$	 &  	-0.438&19.771&1.199&0.988& 39.324&1.62&0.980  \\
\hline    $a_ 7$	 & 0.274&144.34&0.925&0.981& 118.2&0.859&0.980  \\
\hline    $a_ 8$	 &  5.446&665.2& 1.852&0.971& 23.348&0.744&0.900  \\
\hline
\hline    $s_3$  &   0.665&9.726& 0.989&0.911& 5.967  &0.755&0.910  \\
\hline    $s_6$ & 5.821&$exp.t$&$exp.t$&$ 0.970$& 14.12&0.914&0.908\\
\hline    $s_{63}$ &  8.09 &$exp.t$&$exp.t$&$ 0.970$& 14.37&0.764&0.890\\
\hline    $a_4$	 &  2.51&61.69& 1.427  &0.969& 10.932&0.781&0.940\\
\hline    $a_5$	 &  -0.594&2.708& 0.425 &0.895& 3.860&0.612&0.888\\
\hline    $a_{54}$ &  -0.022&13.764&0.844  &0.989& 14.00&0.852&0.989\\
\hline 
\end{tabular}
\caption{ Summary of   fit parameters to NJP data grouped as  for Figs. \ref{fig:HESsubsmall}-\ref{fig:surf}; the  parameters  correspond to the various  formulae discussed in the text, Eqs. (\ref{eq1})-(\ref{ZMeq3});  $q.pw.t$ indicates a strong queen plus power law tail  effect and $exp.t$ indicates a strong exponential tail cut-off, i.e. cases for which the scaling parameters have large error bars;  the regression fit  coefficient R$^2$ is given  for the different cases; data has been rounded up to significant decimals; error bars much below 10\%
}\label{Tablestat3}
 \end{center}
 \end{table}


 \begin{figure}
  \includegraphics[height=9.8cm,width=10.8cm]{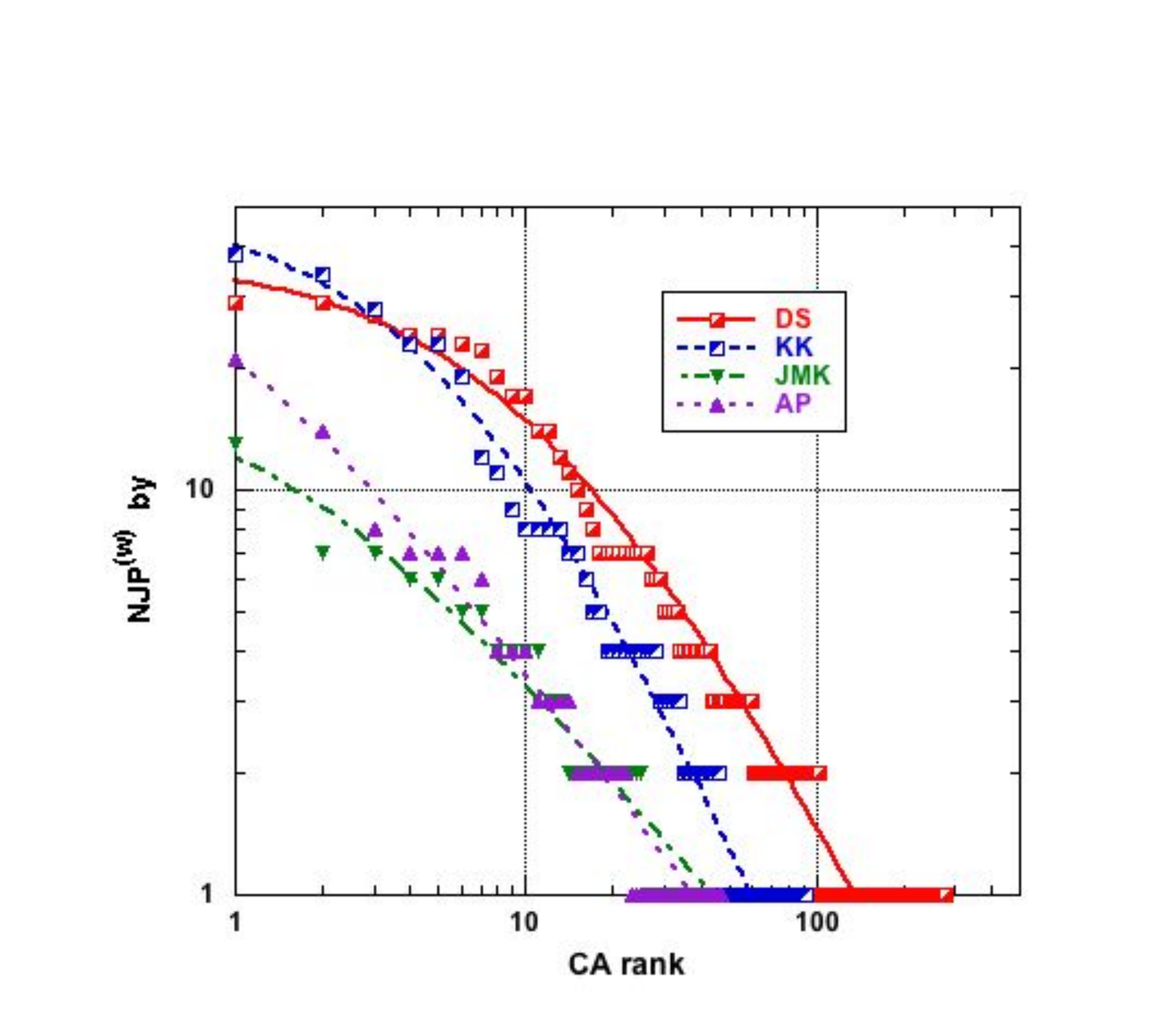}                                                                                               
\caption{  Log-log scale display  of the number of  all joint publications (NJP$^{(w)}$)  by various authors (see insert: DS, KK,  JMK, AP) with co-authors (CA) ranked by decreasing "importance"; best fits (color lines) are shown for the   ZMP law, Eq.(\ref{ZMeq3})
  }
\label{Plot2DSJMKAPloloZMP3}  
\end{figure}

     \begin{figure}
  \includegraphics[height=9.8cm,width=10.8cm]{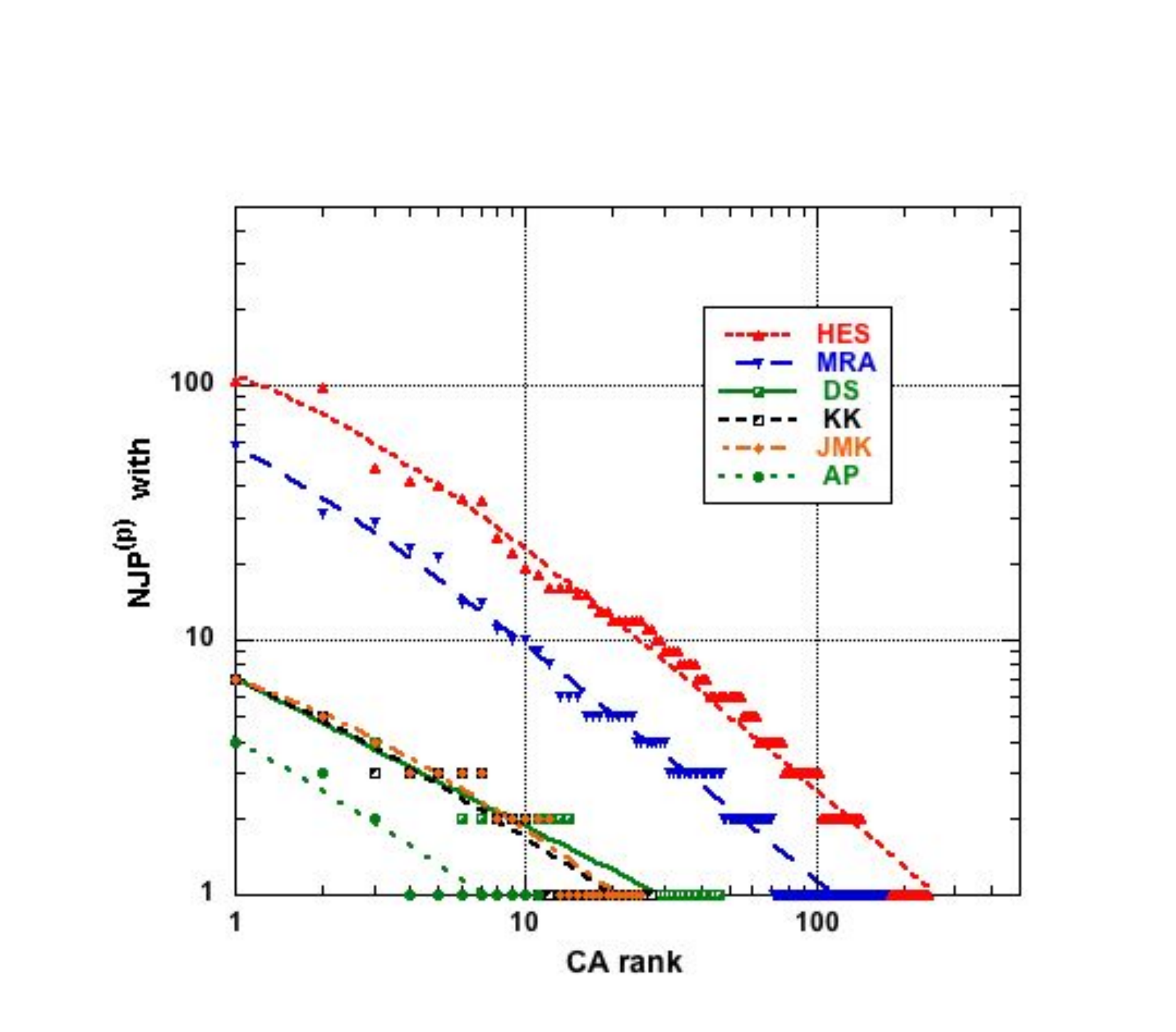}                                                                                               
\caption{  Log-log scale display  of the number of joint publications (NJP$^{(p)}$) in "proceedings", by various authors (see insert) with co-authors ranked by decreasing "importance"; best fits (color lines) are shown for the   ZMP law, Eq.(\ref{ZMeq3}) 
  }
\label{Plot2NJP6lolo} 
\end{figure}
     \begin{figure}
 \centering
  \includegraphics[height=8.8cm,width=10.8cm]{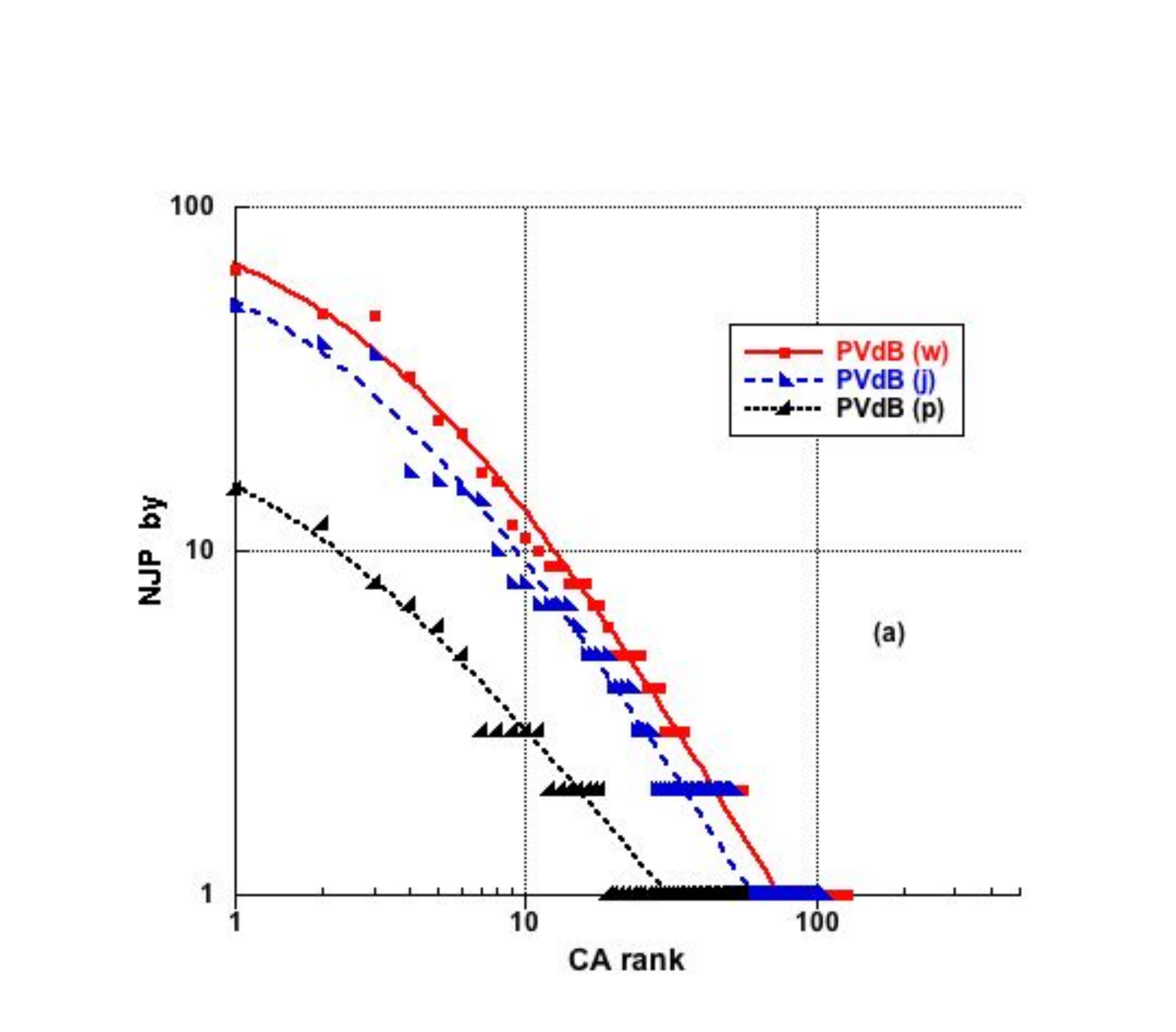}                                                                                              
  \includegraphics[height=8.8cm,width=10.8cm]{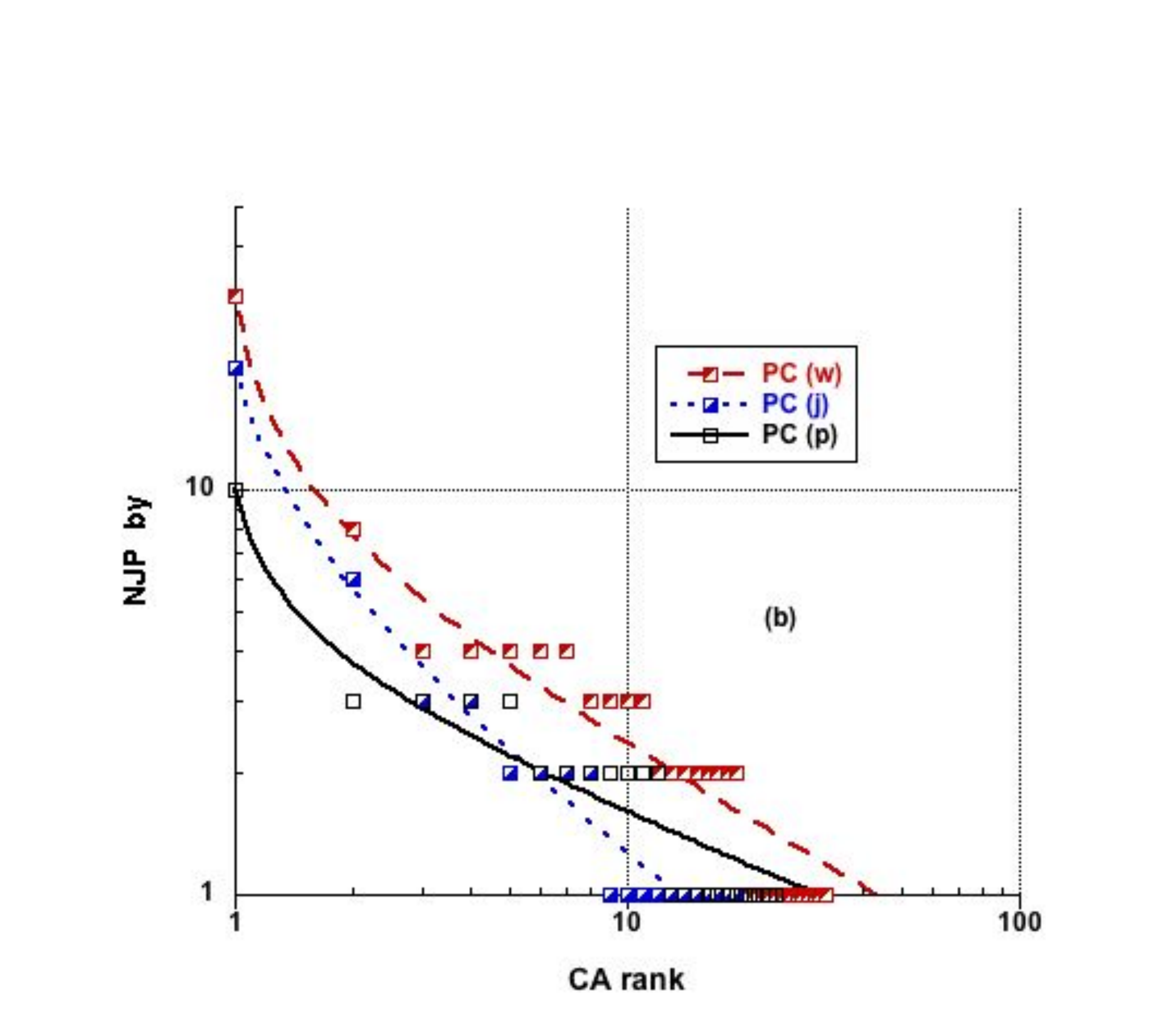}                                                                                               
\caption{  Log-log scale display  of the number of (various types of) joint publications (NJP) by  (a) PVdB and (b) PC, with co-authors (CA) ranked by decreasing "importance"; best fits (color lines) are shown for the   ZMP law, Eq.(\ref{ZMeq3}); note the  "queen  effect" in  (a) and the "king effect" in (b)
  } 
\label{Plot14PCjpjp} 
\end{figure}

 \section{ Amplitudes and exponents. A discussion   } \label{sec:datasetdiscussion}


  After the statistical analysis based on the correlation coefficient, $R^2$,  let the exponent and amplitudes of the empirical laws be examined. first, reconsider  the $whole$ data for the two PIs, HES and MRA,  displayed with the best power law and best ZMP fits on Fig. \ref{fig:MRAHES}. The ZMP is visually much better performing, - the more so if the two lowest ranked data points are removed before a fit to Eqs. (\ref{eq1})-(\ref{ZMeq3}).  The display emphasizes that both HES and MRA have at least two privileged CAs.  This is also going to be  emphasized through the display in  Figs. \ref{Plot2DSJMKAPloloZMP3}-\ref{Plot14PCjpjp}, irrespective of the "very" or "less" plowed research sub-fields.

Observe that such a removal of two outliers, induce a huge change (almost a factor of 3) in the amplitude $b$ and in $\nu$, thereby strongly stressing a queen affect , - due tot he king and vice-roy of HES, and the king of MRA. These effects on $\zeta$ are mild,  though in both cases increase $\zeta$, indicating a faster decay of the empirical law at high rank, - in other words, a weakening of the  role of the rare CAs.

\subsection{Journals}\label{sec:datasetdiscussionwjp}
 
 From Table \ref{Table3peerproceedings}, it appears that  the range of amplitude  can be large and exponents  rather narrow.  The former is due to the wide range in the number of publications of various PIs, but the latter indicates some stability thus a reasonable  guess of the empirical law, - in fact, as should be stressed, whatever the type of publication. Most of the values of $zeta$ are close to 1. The notable exceptions,  pointing to non-universality in contrast indicate the interest of using such an exponent as a measure of the PI  relationship with his/her team of CAs. Obviously, the greater the exponent $\zeta$ the shorter is the team size. The popularity is influenced by the number of rare CAs, thus by a widening of the NJP distribution,thus by a flattening of the  tail, whence by a smaller $\zeta$ exponent.  
 
 Nevertheless, one should disregard  the PIs who are  in fact "not quite PIs", as exemplified by their small number of publications, and the value of $\nu$ being negative, i.e. PC and JM.
 
 In so doing, the most popular PIs  in the presently examined set are (in terms of the $\zeta$ exponent for the whole set of  joint publications) : HES (0.83), JMK (0.87), MRA (0.97) and AP (0.98). It is thus very interesting to notice that  neither DS (1.38) and KK (1.67) nor PVdB (1.44) are "popular".  However, they have large "harem",  - $\nu$ being quite large. This seems quite understandable  due to the activities of the researchers.
 
 Practically, it can be observed that HES presents a king and a vice-roy effect in NPJ$^{(j)}$. Indeed the list of publications of HES shows the high relevance of such two preferred CAS. However,  HES presents rather a queen effect  for NJP$^{(p)}$. Again, this is understandable on the same scientific and popularity ground. He is indeed one of the top masters and most popular in his field. QED.
 
\subsection{Proceedings}\label{sec:datasetdiscussionsub-fields}
  Next, observe  the data when broken into sub-fields as displayed and numerically fitted on 
  Figs. \ref{fig:HESsubsmall}-
\ref{fig:surf},    for   these two  PIs. The displayed data  has been grouped both according to the size of NJP and the  rank range for better visibility; thus in  Fig. \ref{fig:HESsubsmall}, one finds the $i=$ 1, 2, 4; 9, 11 ("less prolific") sub-fields, and in  Fig. \ref{fig:HESsublarge}, the $i=$ 5, 7, 8, 10, 12  ("more prolific") sub-fields for HES.   The case of  sub-fields $i=$ 4 and 5, for MRA, and of  sub-fields $i=$ 3 and 6, for HES, are treated in the Appendix B, and displayed on Fig. \ref{fig:surf},  and compared to the regrouped data into $a_{54}$ and $s_{63}$.      
  
   For better visibility, the classical linear axes plots are limited to the region of interest, NJP $\le$ 40, and NDCA $\le$ 80 or less when convenient,   i.e., where the $m_a$ index can be measured (Ausloos 2013), as suggested by the diagonal line indicating the core threshold. On such linear-linear axes the power law fits do not much indicate deviations. However, on the log-log plots, the ZMP fits are $ll$ obviously better reproducing the empirical data. The queen and  king  effects are often well seen.
   

       It is interesting to observe that $\nu$ is negative in the $a_6$ and $a_5$ cases, and $a_{54}$,  though the latter case is one of merged   fields. This implies a $ strong$  king effect for one CA of  these sub-fields of MRA.
      
      \subsection{Remarks on the king, vice-roy, queen and harem effects}\label{sec:KVRQH}
      
      It is easily understood that the king and  queen effects are at once seen on plots.  Moreover, it has been observed that sometimes the low ranks  contain more than one "equivalent (in terms of NJP)  CA", thus one may  propose the existence  (or definition) of vice-roy(s), on one  hand, and of a harem, on the other hand,  effect. Numerically,  this occurs  respectively when 
      \begin{itemize} \item $\nu \rightarrow $ 0, and more drastically when $\nu\le 0$, $\rightarrow$ -1;
      \item when $\nu$ is large.
      \end{itemize}
      
      Concerning the queen effect, it is  best measured through the ratio $\zeta/\nu$, as seen by taking the derivative of 
      $d\;ln(J)/d\;ln(r)$, i.e.
      \begin{equation}   \frac{dln(J)}{dln(r)} =  -\frac{\zeta\;r}{\nu+r}
      \end{equation}
       which should  be $\preceq$   0, at small $r$.  The harem effect is  quite obvious for DS (w) and DS (j), KK  (w) and KK (j), and PVdB (all types). This is also remarkably true for  the largest $\nu$ cases, i.e. $a_2$ (see also Fig.\ref{fig:MRAsub1-8}), $s_4$ and $s_2$. Those large $\nu$ values indicate that both PIs have published many joint papers with a   short set of CAs, - often the same ones on a large number of publications, and have thus "many queens". In fact, according to the sub-field definitions (Bougrine 2014), these sub-fields with many queens pertain to medical topics, for HES, and experimental work on materials, for MRA. It is understandable that a team effect, with kings and queens,  are to be expected in such (transdisciplinary) domains, thereby stressing the intrinsic  interest of the ZMP law,  whence leading to suggest the following model.
   
     Concerning the king and vice-roy effect,  it has been discussed    through a comparison of HES  with HES* and MRA  with MRA* in Table \ref{TablestatFig1};     the effect  is highly remarkable for JMK (j),  and PC and JM,  on all types of  publications: they have  all a few team co-leaders\footnote{This has been recently examined considering pairs of leading CA through a binary scientific star concept (Ausloos 2014)}. It seems that such effects are strongly constraining the fits, when $\nu\le 0.2$..

         \begin{figure}
\centering
   \includegraphics[height=9.8cm,width=10.8cm]{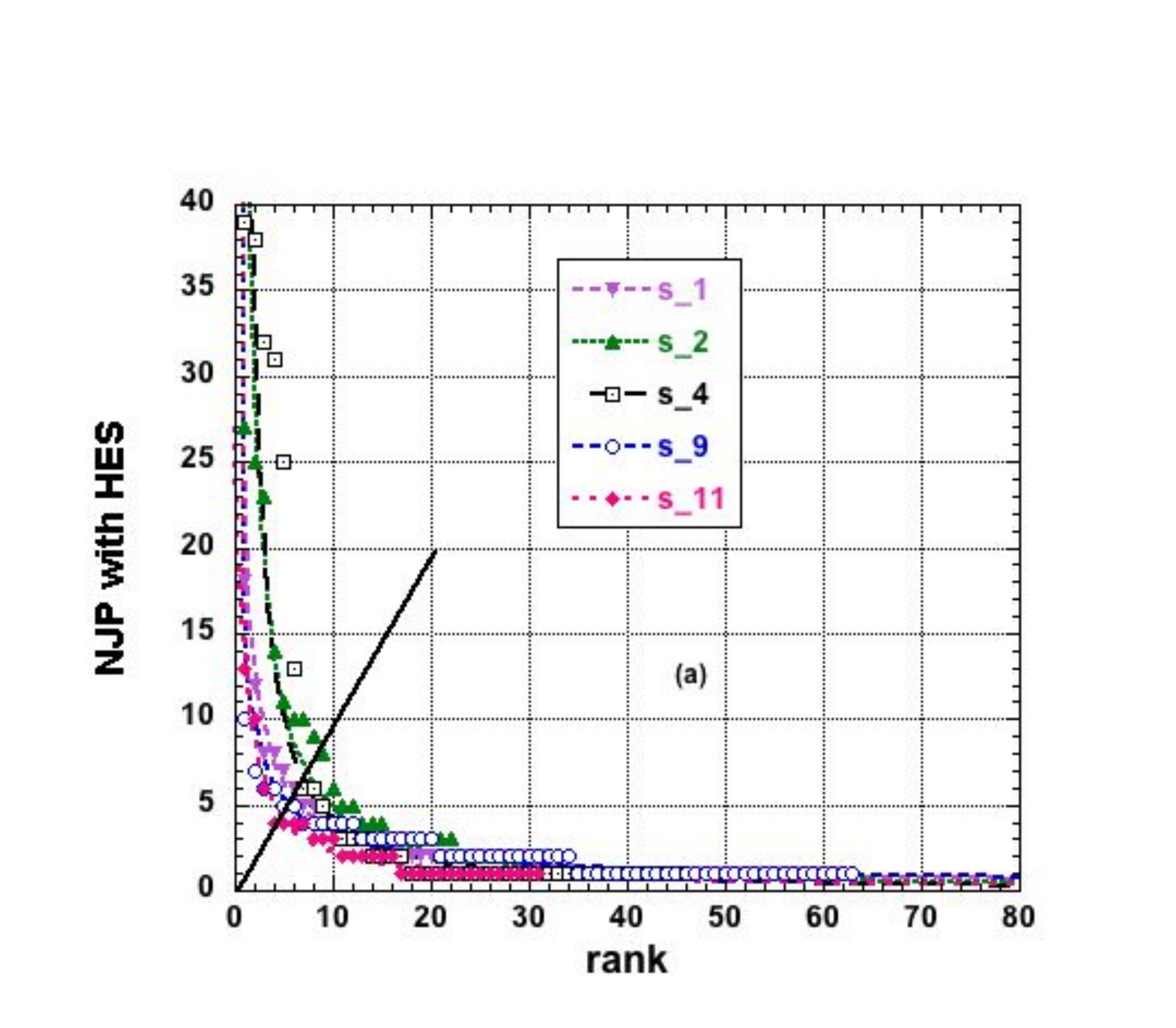}                                                                                              
 \includegraphics[height=9.8cm,width=10.8cm]{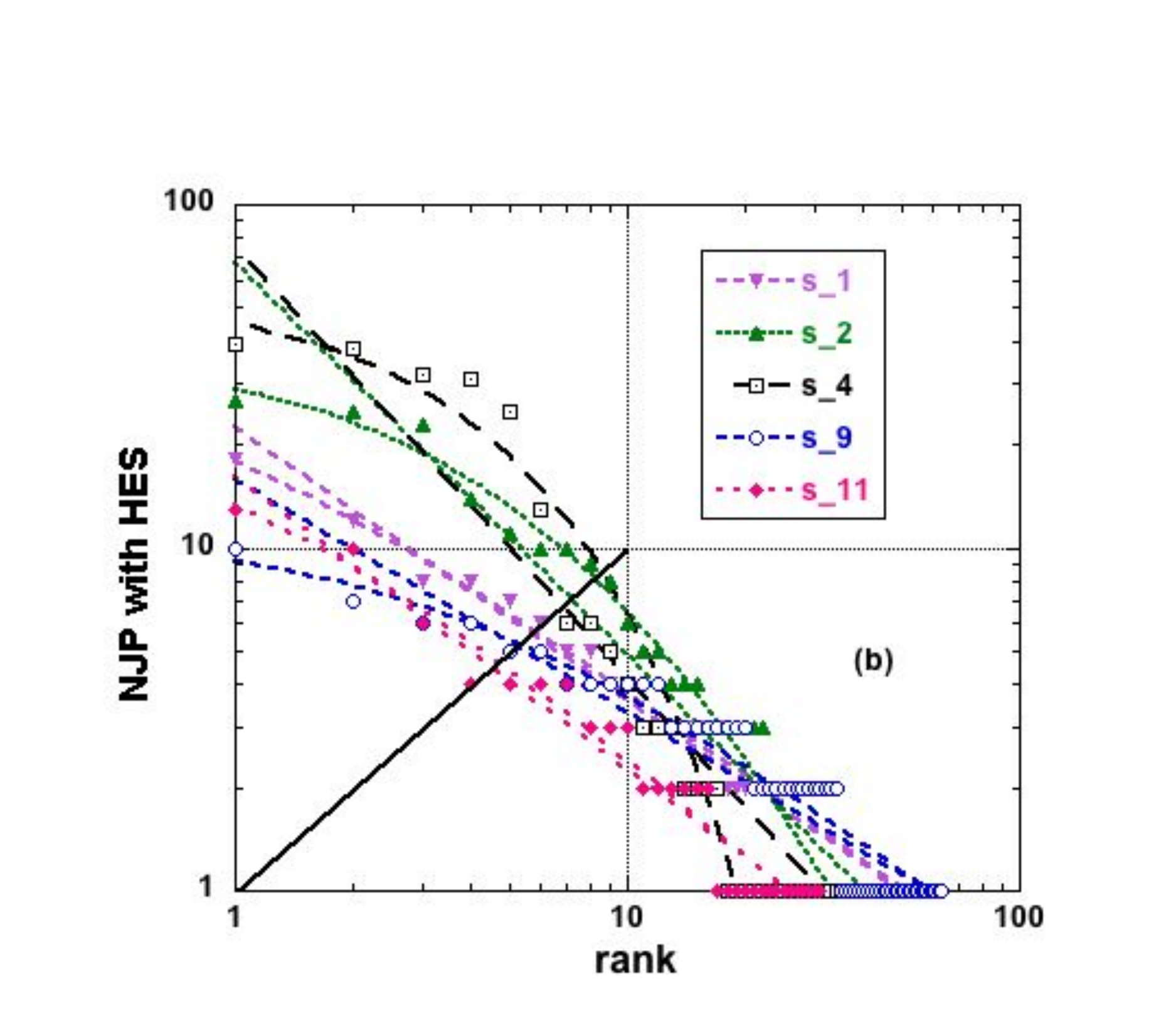}                                                                                                
\caption{  Number of joint publications (NJP) for HES, with co-authors ranked by decreasing importance for  his 5 "less prolific"  sub-fields  $s_i$; (see text for $i=$ 1, 2, 4, 9, 11 definition):  (a)  in the vicinity of the   co-author core measure; (b)  log-log  display  of  (a); best fits, i.e. power law in (a) and ZMP in (b),  are given    for the  overall range; the   co-author core  can be easily deduced from the diagonal line position }
\label{fig:HESsubsmall}
\end{figure}

         \begin{figure}
\centering
 \includegraphics[height=9.8cm,width=10.8cm]{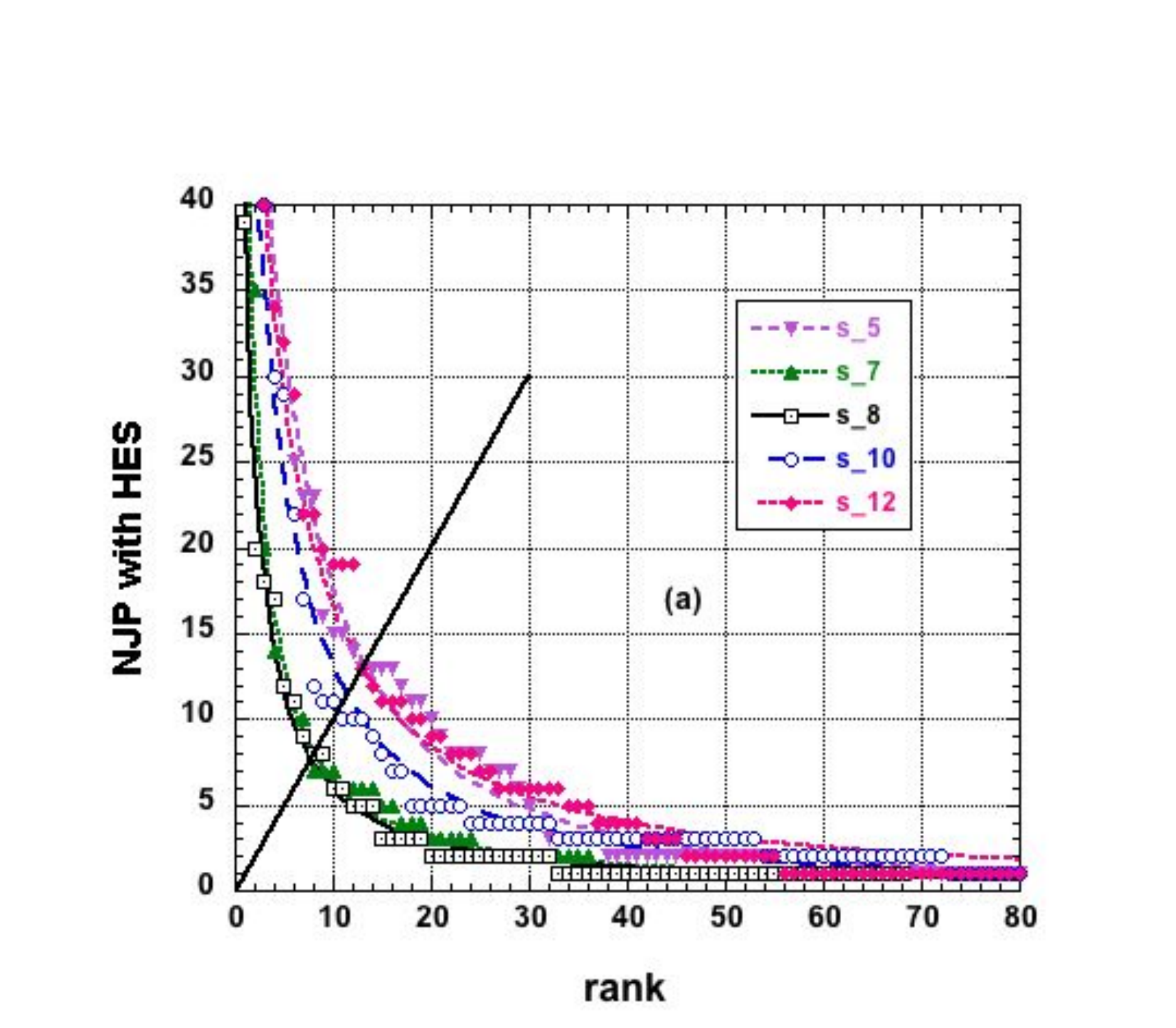}                                                                                              
  \includegraphics[height=9.8cm,width=10.8cm]{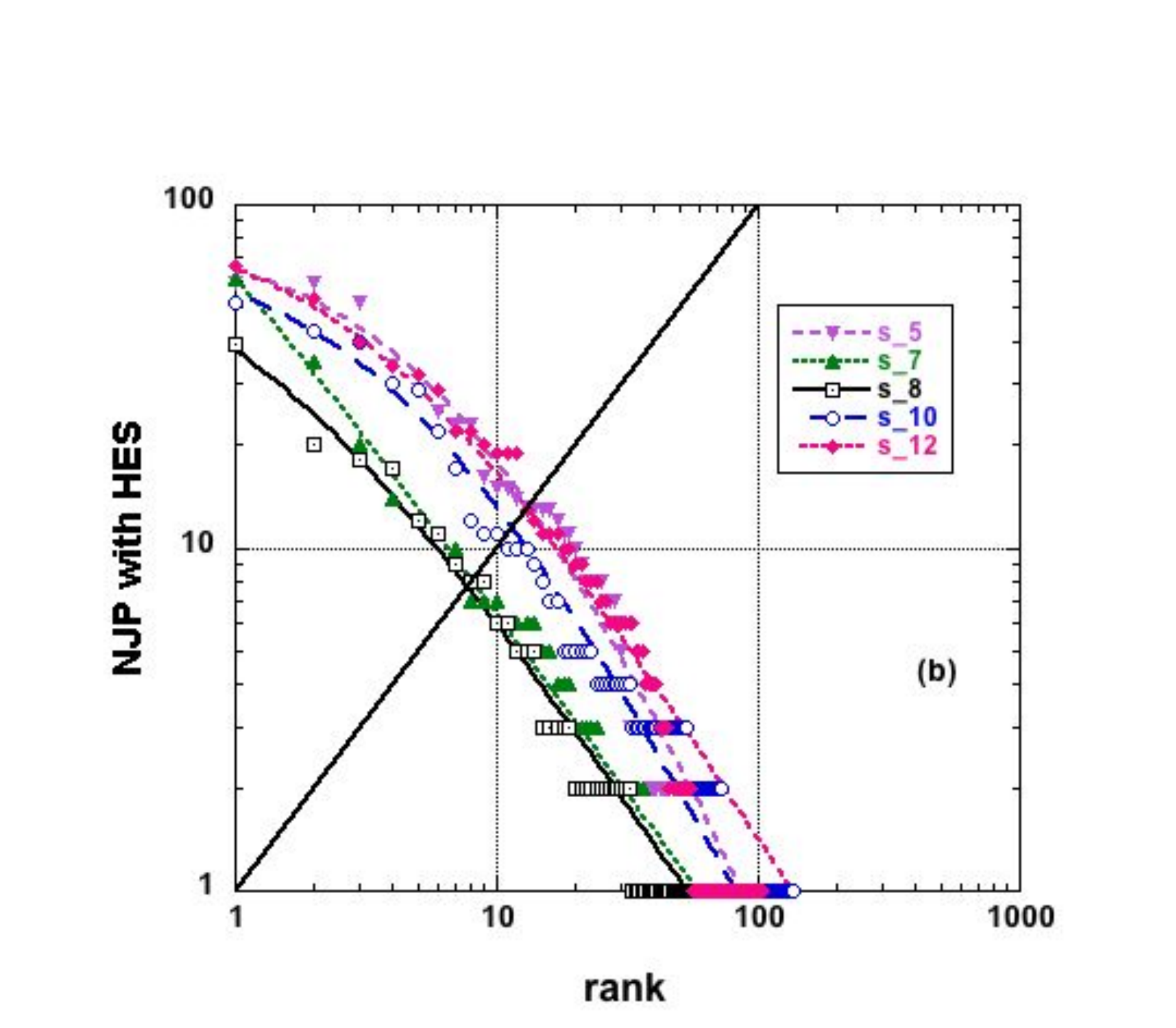}                                                                                              
\caption{   Number of joint publications (NJP) for HES, with co-authors ranked by decreasing importance for the 5 "most prolific" sub-fields  $s_i$;  see text for $i=$ 5, 7, 8, 10,  12  definition;   (a)  in the vicinity of the   co-author core measure (the   co-author core  can be easily deduced from the diagonal line position); (b)  log-log  display  of  (a); best fits, i.e. power law in (a) and ZMP in (b),  are given    for the  overall range }
\label{fig:HESsublarge}
\end{figure} 
 
      \begin{figure}
\centering
  \includegraphics[height=9.8cm,width=10.8cm]{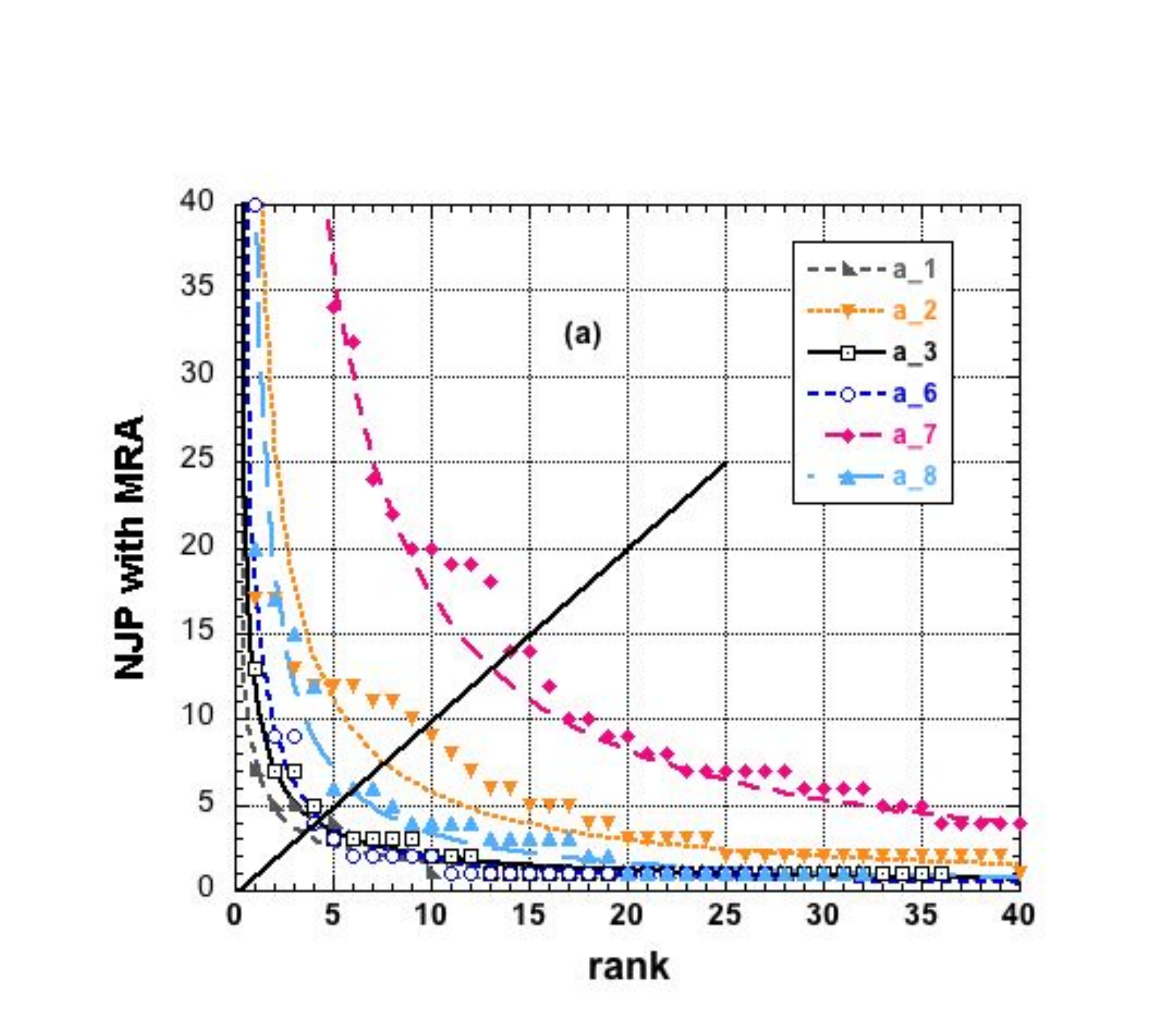}                                                                                              
 \includegraphics[height=9.8cm,width=10.8cm]{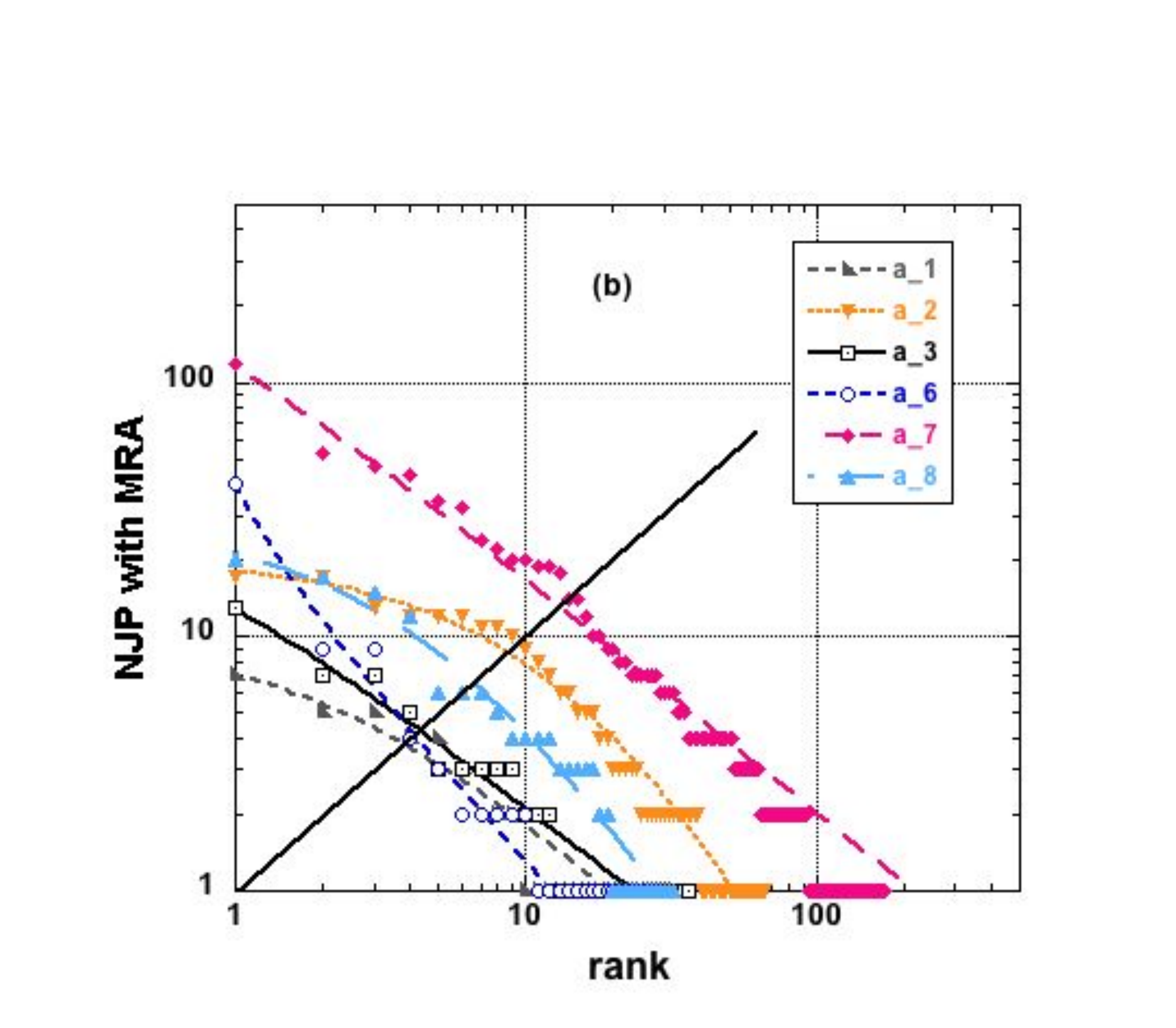}                                                                                              
\caption{  Number of joint publications (NJP) for MRA, with co-authors ranked by decreasing importance for 6 different sub-fields  $a_i$; see text for $i=$ 1, ..., 8 definition;  (a)  in the vicinity of the  co-author core measure; (b)  log-log scale display  of  (a);  the   co-author core  can be easily deduced from the diagonal line position;  the best fits, i.e. power law in (a) and ZMP in (b),  are given    for the  overall range }
\label{fig:MRAsub1-8}
\end{figure}
   
   \begin{figure}
\centering
  \includegraphics[height=10.8cm,width=10.8cm]{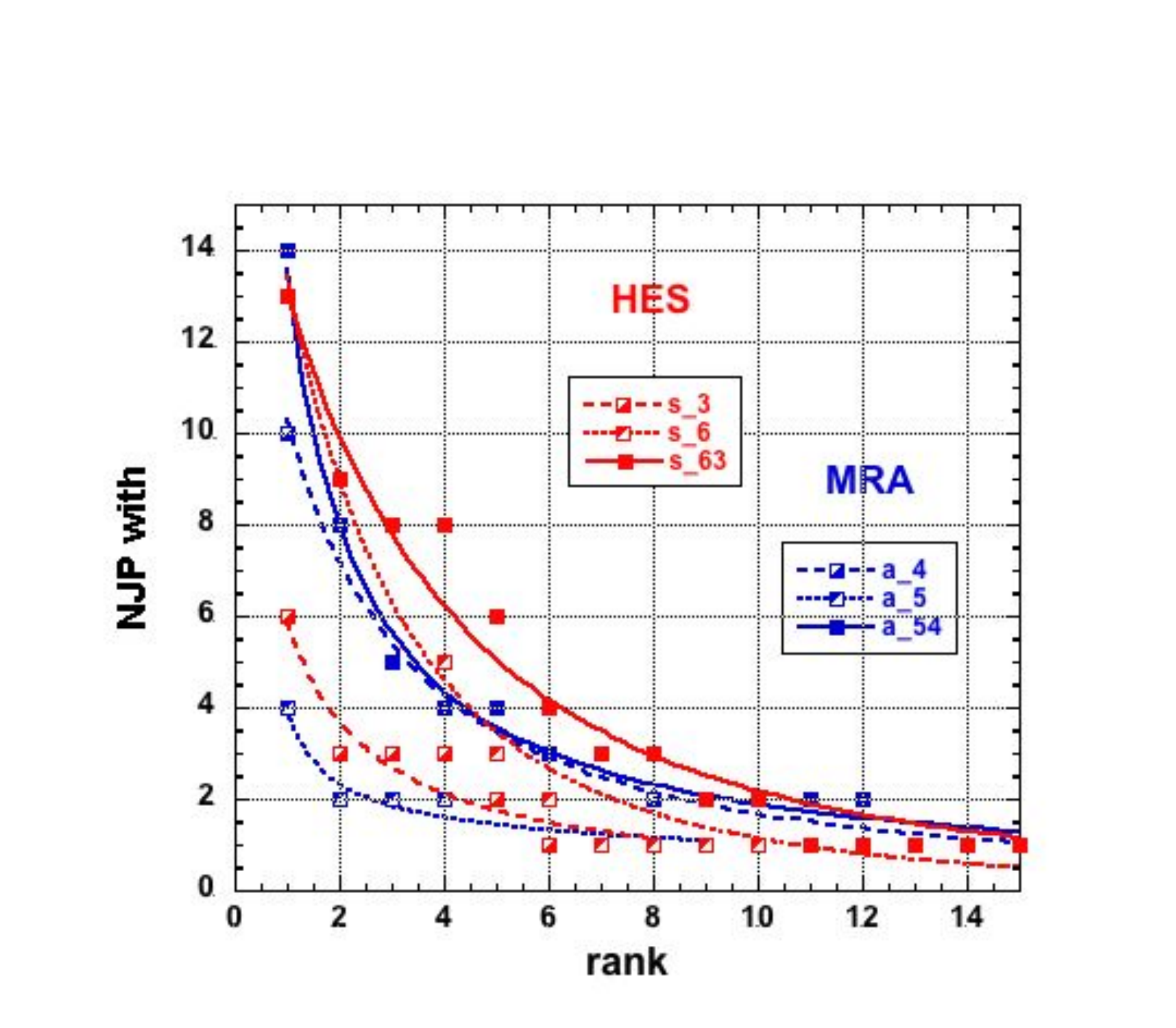}                                                                                              
\caption{    Number of joint publications (NJP) with co-authors ranked by decreasing importance,  in the case of  sub-fields 4 and 5 (for MRA) and 3 and 6 (for HES), and their merging into a sub-field, i.e. $54$ and $63$ respectively; the best ZMP law fits are given  }
\label{fig:surf}
\end{figure}

 \section{Model} \label{sec:model}

 In so doing, it seems reasonable to suggest a model of  team publications, i.e. for a popular PI and his/her CAs,  along the line of Mandelbrot's model for quantitative linguistics  (Mandelbrot 1960, Amati et al. 2002, Manin 2009), in fact similar to  that of city formation (Madden 1956,  Rosen \& Resnick 1980, Glaeser 2008).  
 Recall that  Mandelbrot has  first derived the Zipf's law assuming that the optimization of the  information/cost ratio is resulting from  random character sequences, i.e. the "random typing model".  Mandelbrot's model of texts assumes that the optimal language is one where each sequence of $n$ letters is as frequent as any other (all characters are equiprobable). Thus,   the probability of any given word is   exponential in  length. Whence, each word's  frequency   rank is asymptotically given by a power law.  Nevertheless, note that it has been observed that  the number of distinct words of the same length in a real language,  is far from being exponential in length (Manin 2009).

In fact,  Mandelbrot proposed that   the
language is optimal if it minimizes the average ratio of production cost to information content. This leads to Eq.(\ref{ZMeq3}).  To do so,  Mandelbrot  defined the information content of a word to be the Shannon entropy, i.e. the negative logarithm of the word frequency in the text or language.  

In the same spirit, one can assume to derive  Eq.(\ref{eq1}), in the present case, that the choice of CAs by a PI is at first a stochastic process, - as  proposed by Hsu  \&   Huang (2009). However, as ties developed or not with the PI, either the CAs become $quasi$ happaxes, or maintain a strong tie.  Moreover the CAs attract other CAs (or reject them, for whatever reason), whence organizing a CA utility network, with the PI as hub.

 One can also imagine that the choice of CAs, as time  and research activity progress, has some cost or utility for the PI. He/she recognizes some interest in being associated, and publishing with one or another CA, - whatever the profound reasons. Such an  optimization principle allows  to demonstrate how the "optimal state" can be achieved as a result of research evolution. The $\alpha$ exponent is close to 1, as in the case of  language and   of city population size optimization, i.e.  
 a desirable situation in which the  forces of concentration of coworkers balance those of  competition by other external teams, called "decentralization" in city formation  (Zipf 1949, Madden 1958, Glaeser 2008). In co-authorship popularity measures, as well, the exponent $\alpha$ (and $\mu$ or $\zeta$) can be interpreted as being close to 1, because of the "same"  type of balance between attraction and rejection of coworkers by the PI his/her king and  queens. Thus, one considers detachment, beside preferential attachment,  which by itself would lead to  a much heavier tail exponent $\zeta$.

Thus, by similarity, along such a line of reasoning, consider  the   joint publication cost per CA  to be  
  \begin{equation}\label{Mcost}
 C= \sum p_r\; c_r
 \end{equation}
 such that the "cost of a publication" with the CA at  rank $r$ is $c_r$, and $p_r$ is the normalized value of $J_r$, - normalization with respect to the whole histogram surface\footnote{For simplicity  of the writing, $r$ is taken as a continuous variable though it is manifestly a positive integer only.} $J(r)$. Let the average information entropy per CA be also assumed to read \begin{equation}\label{CAentropy}
 H= -\sum p_r\; log_2 (p_r).
 \end{equation}
 It is easy, e.g. by the Lagrange multiplier method  (Manin 2009),  to show   
  that  the rank-frequency distribution $p_r$  will minimize the cost ratio $\hat C\equiv C/H$ and will have a ZMP law form if one assumes the individual cost function to be
  \begin{equation}\label{cr}   c_r = c_0\: log_2 (\nu+r), 
   \end{equation}
 which can be negative or positive.  The cost is of course high when $r$ is large. 
 Note that if $\nu=0$, one recovers the Eq.(1) power law.  Eq.(6) assumption (or ansatz)  
  can be accepted if one considers that any research team  has some hierarchical structure with a set of CA at different  levels, the set increasing with the "distance" ($\sim\;r$) from the PI.  For example, the CAs   can be grouped according to a visual rule\footnote{Benguigui and  Blumenfeld-Lieberthal (2011)  are perfectly right :  (text adapted, but resulting from a $quasi$ exact quotation)    in order to be able to decide if Eq. (1)  is (and Eqs. (2-3) are)  verified or not,
one has to fit the data to several functions and compare the results, using the same
criterion. Naturally, it is not realistic to expect each [ $J(r)$ ]   would be fitted to
numerous formulas; thus, we ( $\simeq I$) propose to use a visual inspection in order to help
decide which formulas might represent the data correctly.  ...  we  ( $\simeq I$)  trust the human mind and believe that a visual inspection can indeed give essential information; particularly it helps deciding if the studied system
is homogeneous or not  ...  a simple visual inspection ... shows
that the system  (...)  is not homogeneous. It can be divided into ... 
subsystems. This (...)  emphasizes the need for a visual inspection of the rank-size relation of the real data on log-log scales. This gives the possibility to see (in
the simple meaning of the word, see with the eye) if the points may be fitted with
some mathematical function (not necessarily a straight line).}, based on the behavior of $J(r)$ examined on linear or log-log plots,  
  Fig.\ref {fig:HESgroups}. Practically, this is supplemented by  trial and errors  of  $local$ fits in moving windows of different sizes;  the groups are  deduced and confirmed, choosing them  from the best R$^2$ among various  realistic attempts.  More complicated functions (Popescu et al. 2010, Voloshynovska 2011) could be used, but a detailed parameter interpretation for such function combinations is  a challenging task and requests  further study, much outside this report. 
  
  Nevertheless, the CAs  can  be seen as forming "circles" around the PI, as in e.g. the HES case, Fig. \ref{fig:Plot16HEScircles}. Note that this can be considered as  the  true PI community, - a visible college, since  the CAs of CAs not involving the PI, have been removed from the start,  by data construction. In this cluster, the PI has {\it necessarily at least one link} with some CA.

  The cost of retrieving the CA  at the $r$-th level can be, in a first approximation, assumed to be related  to the NJP $J(r)$ of the CA, or to the  rank,  i.e. $ log_2(r)$. Therefore, the parameter $\nu$ can be understood to indicate that several "far ranked" CAs have to  deal with some other "preferred CAs" of the PI in the joint publication process.  
 
          \begin{figure}
\centering
 \includegraphics[height=9.8cm,width=10.8cm]{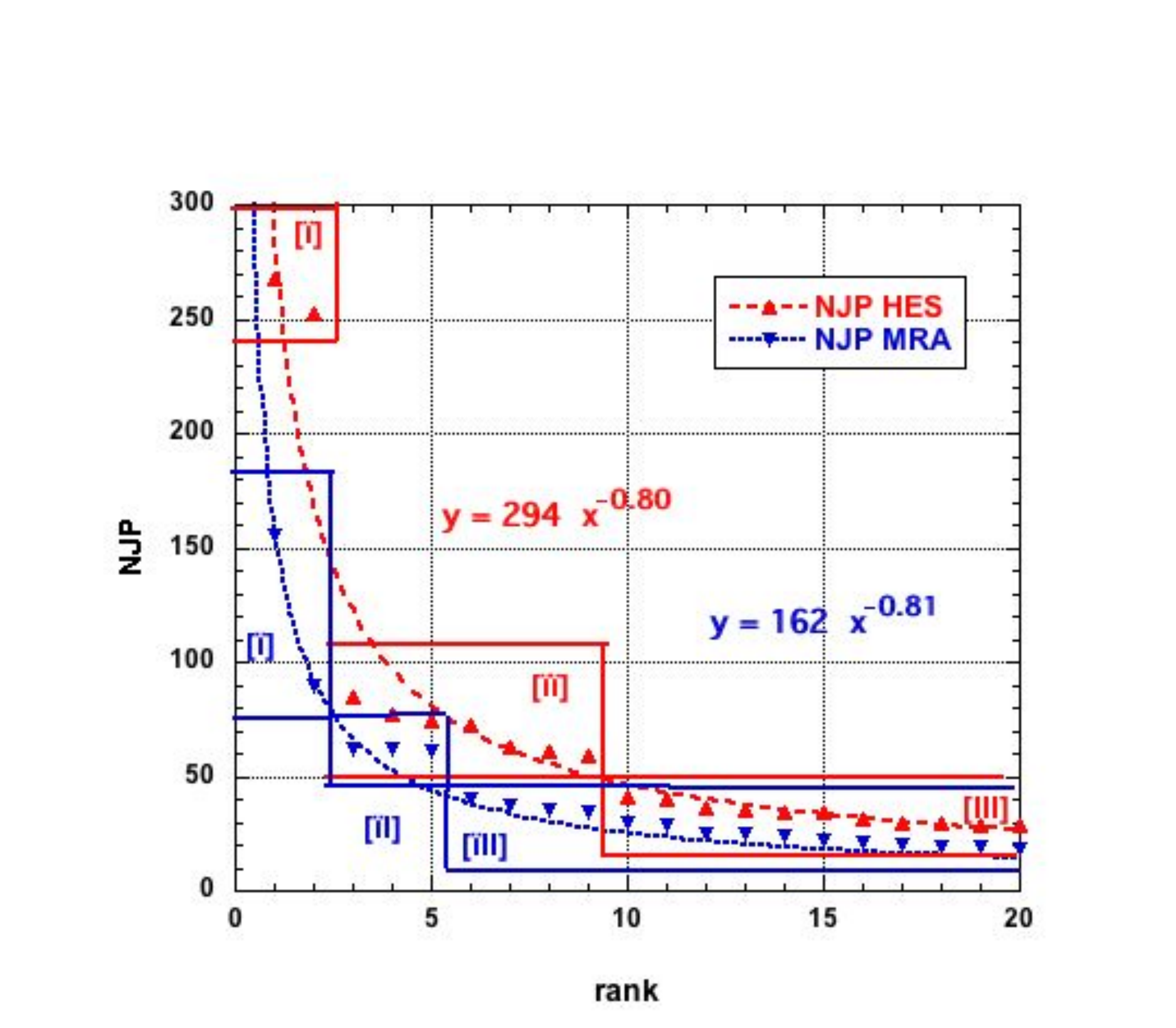}                                                                                              
  \includegraphics[height=9.8cm,width=10.8cm]{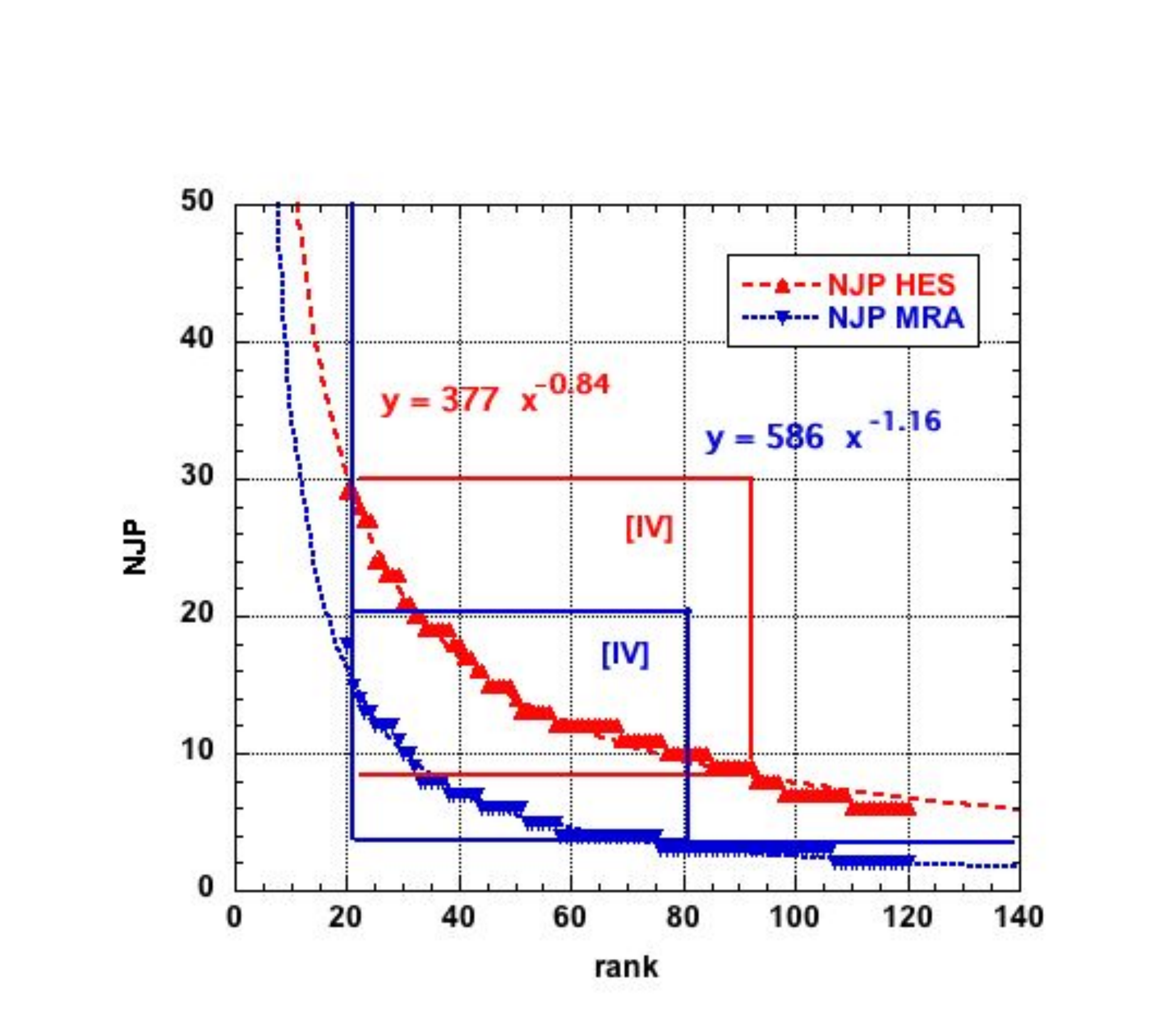}                                                                                              
\caption{   Total number of joint publications (NJP) for HES and MRA, with co-authors ranked by decreasing importance allowing some vizualization of major groups of CAs;   (a)  for low ranks, groups I, II, and III) and (b)  for group IV; the best $local $ power law fits are given as a guide to the eye}
\label{fig:HESgroups}
\end{figure} 

         \begin{figure}
\centering
 \includegraphics[height=9.8cm,width=10.8cm]{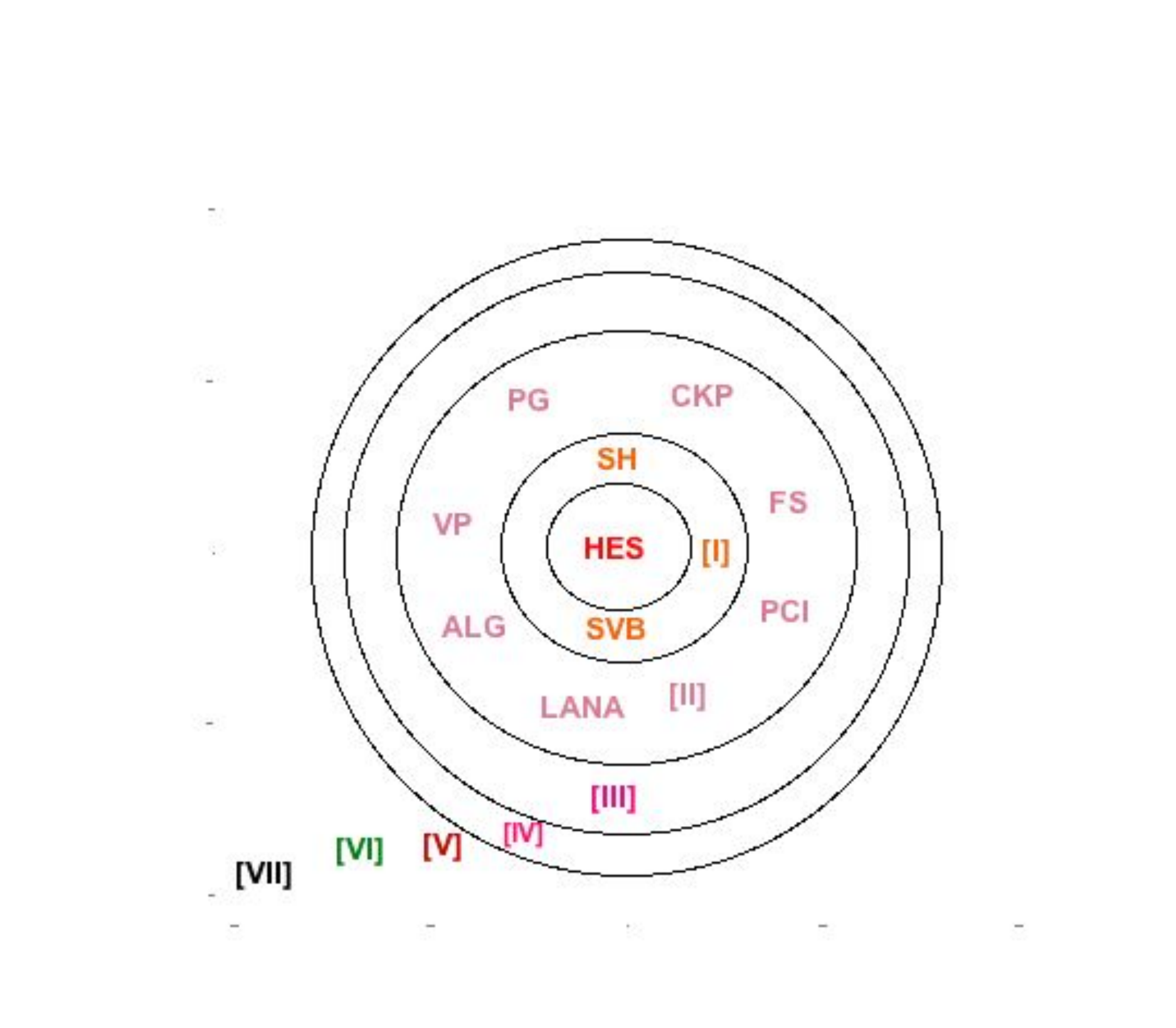}                                                                                               
\caption{   CAs of HES    distributed in the I, II,   III, and  IV  group according to their NJP}
\label{fig:Plot16HEScircles}
\end{figure} 
     
  \section{Conclusions}
\label{sec:conclusions}

    Most prolific scientists have joint publications on different subjects. Thus, co-authors might be specific to some research sub-field  of a PI. It is thus if interest to examine, for such teams and leaders, whether the $j(r)$ relationship (Ausloos 2013)  is obeyed when the research publications pertain to different sub-fields. 
 Bougrine (2014) has   broken   the  publication list  of two such authors into sub-fields.  

Observed irregularities  have been thought to be due to different causes: publication inflation, proceedings counting,  co-authorship inflation,  and/or promotion of  team size visibility. 
    Other "intrinsic causes" might have  arisen from the large productivity of the group based on a high turnover of young researchers, with $r>> 1$,  as well as a steady contribution from stable  partners, with $r\simeq1$.

     Here, deviations have been searched through sub-field considerations. By using the Zip-Mandelbrot-Pareto function for the distribution $J(r)$, thus having one more parameter ($\nu)$, it has been observed that the statistics on one hand is much improved. On the other hand, the introduction of the extra parameter, and the observation of its range of values, has suggested  a model based on a co-author cost scheme, similar to the models of  language usage and city population size distributions.      One must admit that the model is a rather  common sense model to anyone having worked with co-authors, in large or small teams.

 However, the  analysis   has indicated the sensitivity of the sub-field notion, on one hand, and of the co-author distribution, on the other hand. Moreover the notion of cost and utility of co-authorship are introduced,  but demand further  elaboration; some work will be useful along such lines,  for better quantification also. 
 
 Notice that the cost can be positive or negative according to the value of $\nu$; see  Eq. (\ref{cr}). This observation  can be considered to be already useful in order to imagine that one can be  introducing selection and rewarding policies in the career of members of teams,  along the  co-author core measure.  Technically, one could thus measure the relevant strength of a research group centered on some leader by combining the $h$-index and the $m_a$ measure in  a three dimensional ($J$,  $q$, $r$) space taking into account the quotations $q$ of co-authored papers.

  \bigskip  \bigskip 

{\bf Acknowledgements} Thanks to J. Miskiewicz and H. Bougrine for private communications on   their respective work, comments prior to manuscript submission and making available the relevant publication list  data mentioned in  the text.  I warmly  thank all colleagues who have kindly provided relevant data. This paper is part of scientific activities in COST Action TD1210.
 \bigskip

  \begin{table}  \begin{center} 
 \begin{tabular}{|l|c|c|c|c| c|c|c|c|c|c| c|c|c|c|c|c|   }
   \hline
  \multicolumn{1}{|c|}{ }&\multicolumn{3}{|c|}{4 param. ZMP, Eq.(3)}&\multicolumn{3}{|c|}{ 3 param. ZMP, Eq.(2)}  &   \\ 
\hline
 NJP  & $\eta$& $\lambda$& $c$&$\nu$& $b$& $\mu\equiv\zeta$&R$^2$    \\
   \hline    
\hline 
HES  (w)		&0.177&0.376&165.9 	&0.10& 335.0&0.835&0.916 	 \\
 HES$^{*}$ (w)	&1.741&0.236&230.6 	&7.384&1128.0&1.098&0.987  \\
MRA	 (w)	&0.523&0.558&161.7	&0.876&276.25&0.975&0.988   \\
MRA$^{*}$ (w)	&1.547&0.266&206.7	&4.149& 723.3&1.203&0.988 	   \\
  \hline
\end{tabular}
\caption{ Summary of   fit parameters to NJP data corresponding to Fig.\ref{fig:MRAHES}; the  parameters  correspond to the various  formulae discussed in the text, Eqs.(\ref{ZMeq3})-(\ref{ZMeq4});    HES$^{*}$ and  MRA$^{*}$   indicate that the two lowest rank data points corresponding to a strong king and vice-roy  effect have been removed before  fit attempts;  the regression fit  coefficient R$^2$ is given  for the different cases; data has been rounded up to significant decimals
}\label{TablestatFig1}
 \end{center}
 \end{table}

\begin{table}  \begin{center} 
 \begin{tabular}{|c|c|c|c|c| c|c|c|c|c|c| c|c|c|c|c|c|   }
   \hline
  \multicolumn{1}{|c|}{ }&\multicolumn{5}{|c|}{4 param. ZMP, Eq.(3)}    \\ 
\hline
 NJP &  $\eta$& $\lambda$& $c$ & $\mu $&R$^2$    \\
  \hline 
\hline    $s_1$	 & 2.957&4.442&96.058 &0.838&0.988 \\
\hline    $s_2$	 & 1.715&0.181&134.08&  2.074&0.978 \\
\hline    $s_4$	 & 1.176&0.021&503.44&   $q.pw.t$& 0.940 \\
\hline    $s_9$	 & 246.4&64.59&1262.5& 0.859&0.961 \\
\hline    $s_{11}$ & 2.989&3.126&79.342&  0.989&0.974 \\ 
\hline
\hline    $s_5$	 & 1.401&0.188&144.63&	 $1.68$&0.979   \\
\hline    $s_7$	 & 0.257&1.308& 98.392&  1.054&0.994   \\
\hline    $s_8$	 & 1.346&1.148&105.54& 1.128&0.982   \\
\hline    $s_{10}$	 & 1.227&0.291&99.365& 1.427&0.982  \\
\hline    $s_{12}$	 & 1.527&0.264&154.89& $1.19$&0.987   \\
\hline 
\hline
\hline    $a_ 1$	 & 1.668&0.644&17.642&  1.088&0.939   \\
\hline    $a_ 2$	 & 2.048&0.058&241.0   &$q.pw.t$&0.980  \\
\hline    $a_ 3$	 & 1.868&5.118&73.331&  0.893&0.971   \\
\hline    $a_ 6$	 & -2.013&2.769&31.154& 1.199&0.988   \\
\hline    $a_ 7$	 & 0.219&0.797&117.1& 0.925&0.981   \\
\hline    $a_ 8$	 & 1.96&0.292&113.9&  1.852&0.971   \\
\hline
\hline    $s_3$  & 0.734&1.098&10.69 &   0.989&0.911  \\
\hline    $s_6$ & 0.964&$exp.t$&15.39 &  $exp.t$&$ 0.970$ \\
\hline    $s_{63}$ & 1.016&0.056&17.68& $exp.t$&$ 0.970$ \\
\hline    $a_4$	 & 2.137&0.851&49.00&   1.427  &0.969 \\
\hline    $a_5$	 & -2.963&4.991&5.374&  0.425 &0.895 \\
\hline    $a_{54}$ & -0.075&3.479&39.44  &  0.844  &0.989 \\
\hline 

\end{tabular}
\caption{ Summary of   fit parameters to NJP data grouped as  for Figs. \ref{fig:HESsubsmall}-\ref{fig:surf}, according to different sub-fields of HES ($s_i$) and MRA ($a_i$); the  fit parameter values  correspond to  the 4-ZMP form, Eq. (\ref{ZMeq4});   $q.pw.t$ indicates the presence of both a strong queen and power law tail  effect, while $exp.t$ indicates a strong exponential tail cut-off, i.e. cases for which the scaling parameters have large error bars;  the regression fit  coefficient R$^2$ is given  for the different cases; data has been rounded up to significant decimals; error bars $\le$ 10\%
}\label{TablestatZMP4sa}
 \end{center}
 \end{table}  
 
  \bigskip
\newpage
\section*{Appendix A. ZMP fits  with 3 or 4 free parameters }  \label{AppZMP4}
 
 Using  the 3-parameter free ZMP   function, Eq.(\ref{ZMeq4}), for data  fitting is much more troublesome than fitting  with the Zipf hyperbolic law (Fairthorne 1969, Haitun 1982,  Iszak 2006).  Thus, a variant of the ZMP law, i.e. the 4-parameter  relation
Eq.(\ref{ZMeq4}) is sometimes proposed, since it allows for one more scaling parameter. It is  often observed that the 4-ZMP has some advantage with respect to the 3-ZMP, from the point of view of  the stability of the solutions of the non linear system of equations for the fit parameters. This is interpreted as due to the fact that the numerical values of the other parameters ($\mu$,  $\eta$, $\lambda$, and the more so $c$) fall into more compact ranges. For examples, compare  the amplitudes  $c$  and $b$ for $s_2$ and $s_4$, respectively,  in Tables \ref{Tablestat3} and Table \ref{TablestatZMP4sa} for the 4-ZMP and 3-ZMP fits.

However nothing drastic has been found in the present cases,  as seen from Tables  \ref{TablestatFig1}-\ref{TablestatApp}. Moreover,  the  meaning of $\nu$, in the 3-ZMP case  seems more easily interpretable than the $\eta$ and $\lambda$  values  in the 4-ZMP.

  It should be emphasized that  the  R$^2$ values are identical, up to the third decimal, for the 3- and 4-ZMP parameter law fits,  see  Table  \ref{TablestatZMP4sa}, except for $s_6$ and subsequently $s_{63}$, nevertheless  found close to each other, as likely due to an $exp.t$ behavior pointing to a strong exponential tail cut-off,  - in which cases the empirical laws can be hardly expected to hold.   
  Thus,  it is observed that $\mu \equiv \zeta$ in all   cases, - i.e. the relevant conclusion.

\section*{Appendix B.   On merging sub-fields}\label{AppAmergsub-fields}

\bigskip In order to investigate the effect of reduced size of data in  considering sub-fields,  Bougrine (2014)  merged 2 sub-fields into a single one, both in the case of MRA and HES.  
  For comparison, and completeness,   ZMP and power law  fits have been made on $a_4$ and $a_5$ merged into $a_{54}$ on one hand, and on $s_3$ and $s_6$ merged into $s_{65}$ on the other hand. The parameters resulting from the fits are given in Table \ref{Tablestat3}. The fits are displayed in Fig. \ref{fig:surf}. In such cases, with not many data points, the co-author core is low, and the effect of  many CAs at rank $r\ge 4$ or 6 respectively is rather important.  Thus, the instability of the fits with respect to initial conditions  is due to the presence of a strong  exponential  cut-off  superposed on the power law tail.

These features indicate the sensitivity of the sub-field definition, on one hand, and of the co-author distribution, on the other hand.

\bigskip

\begin{table} 
 \begin{tabular}{|c|c|c|c|c| c|c|c|c|c|c| c|c|c|c|c|c|   }
   \hline
  \multicolumn{1}{|c|}{ }&\multicolumn{3}{|c|}{4 param. ZMP}&&\multicolumn{3}{|c|}{ 3 param. ZMP}&&&\multicolumn{3}{|c|}{ power law}  \\ 
\hline
  CDF&  $\eta$& $\lambda$& $c$&&$\nu$& $b$& $\mu\equiv\zeta$&R$^2$&&$a$&$\alpha$&R$^2$    \\
   \hline
\hline 
 $a_2$&1.117	&0.735&1.154& 	&1.515	& 1.626	&1.120	&0.970  	&&0.603	&0.943	& 0.960  \\
  $a_7$&7.127&4.408&7.759 &		&1.597&1.584&1.143&0.996& &0.898 &1.032&0.932    \\
 \hline
\end{tabular}
\caption{ Summary of  fit parameter values to $a_2$ and $a_7$ frequency-size  cumulative distribution function (CDF) data  with  notations   explained in the text, - corresponding to Figs. 6 and 7??? respectively;  the  parameters  correspond to the various formulae discussed in the text, Eqs.(1)-(3);  the regression fit  coefficient R$^2$ is  given  for the different cases
}\label{TablestatApp}
 \end{table}  

     \begin{figure}
\centering
  \includegraphics[height=14.8cm,width=14.8cm]{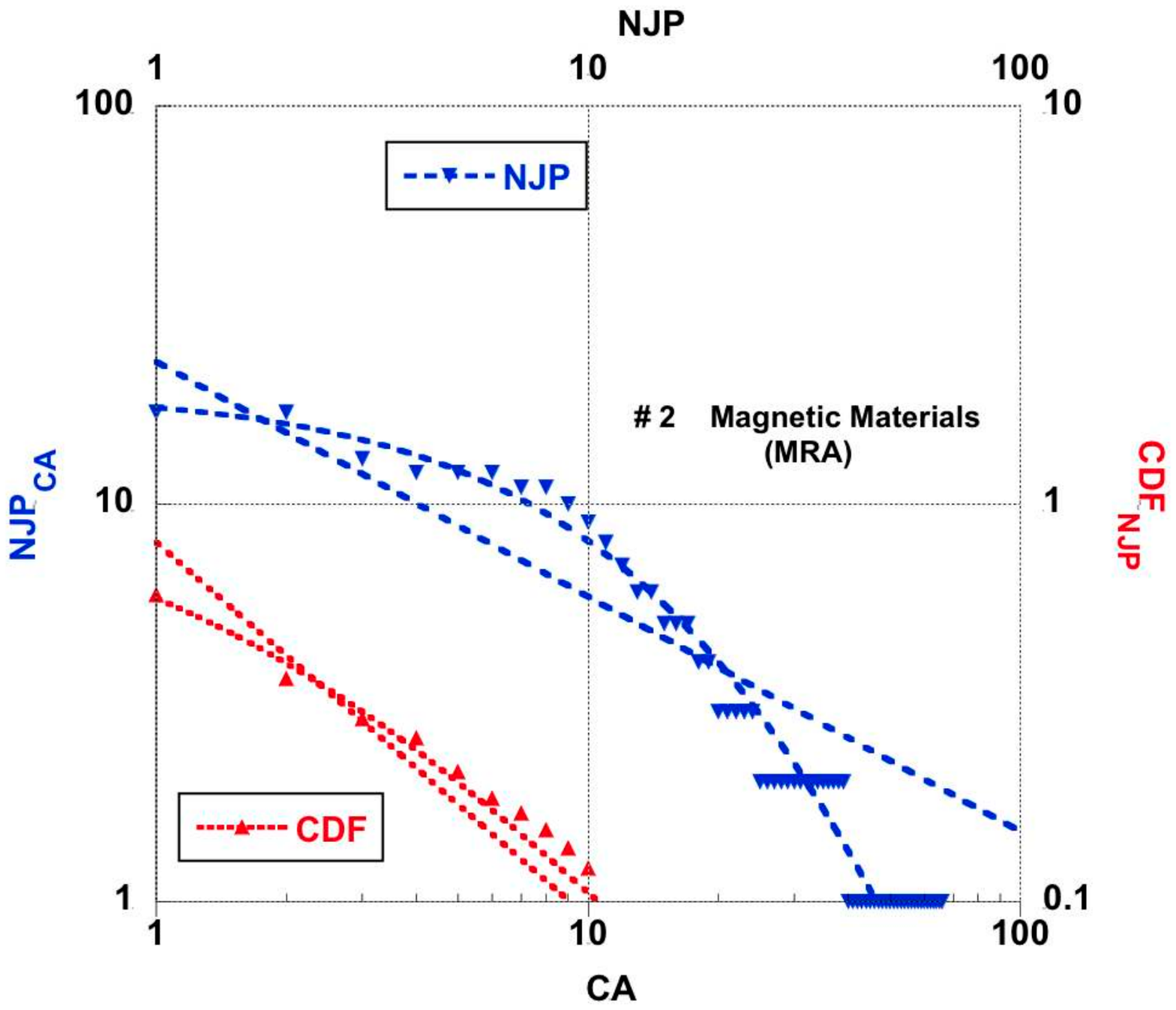}                                                                                              
\caption{   Log-log scale display  of the number of joint publications (NJP)  on  magnetic materials,   for  MRA, with co-authors, ranked by decreasing importance, and of the corresponding       frequency-size  cumulative distribution function     (CDF);   best fits, over the   whole data range are shown for the power law and ZMP law  
 }
\label{fig:MRAmag}
\end{figure}

     \begin{figure}
\centering
  \includegraphics[height=14.8cm,width=14.8cm]{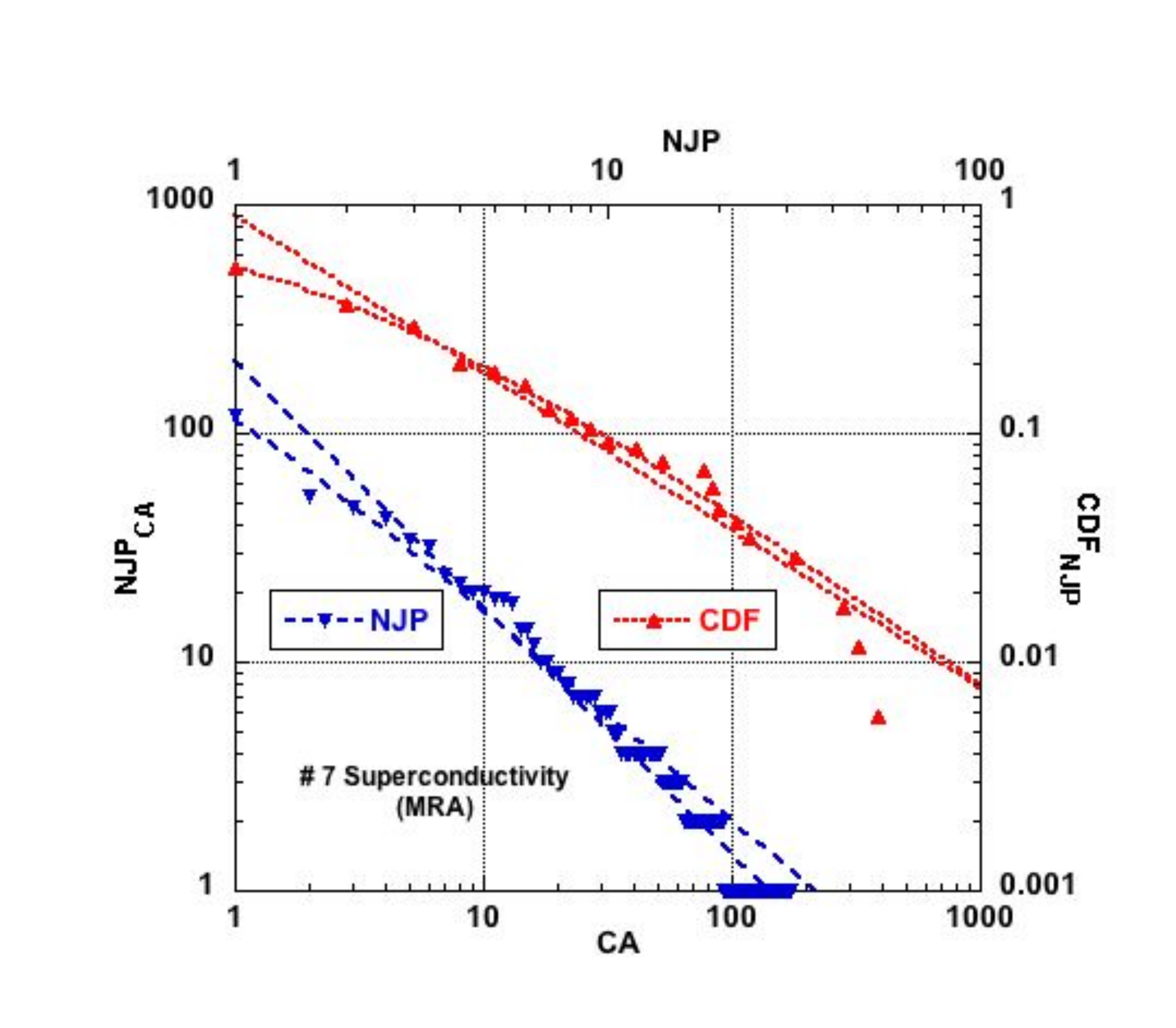}                                                                                              
\caption{    Log-log scale display  of the number of joint publications (NJP)  on  superconductivity,  for  MRA, with co-authors, ranked by decreasing importance, and of the corresponding       frequency-size  cumulative distribution function     (CDF); best fits, over the   whole data range are shown for the power law and ZMP law  
  }
\label{fig:MRAsup}
\end{figure}

\section*{Appendix C.   On cumulative distribution functions (CDF)}\label{AppBforCDF} 
\bigskip

In  Informetrics, one prefers to fit   empirical data to some size-frequency  functional form using a maximum likelihood fit,  rather than  making a least squares fit   for the rank-frequency distribution.   Indeed,    
 one can also ask, as did Pareto (1896),  how many times one can find 
  an  "event" greater than some size $y$, i.e. study  the {\it size-frequency  relationship}. Pareto  found out that  the   cumulative distribution function (CDF) of such events  follows an inverse power of $y$, or  in other words,  $P\;[Y>y] \sim y^{-\kappa}$.   Thus, the (number  or) frequency $f$ of such events   of size $y$, (also) follows an inverse power of $y$. 

Thus, for illustration, ZMP and power law fits have been made on two  of MRA major sub-fields, i.e. $a_2$ and $a_7$. A log-log scale display  of the number of joint publications (NJP)  with co-authors ranked by decreasing importance and the corresponding CDF     are  shown in Figs. \ref{fig:MRAmag}-\ref{fig:MRAsup}.  Both the power law and ZMP law fits are shown  for the  all $r$ range.  Note that the NJP data  and fits are those  seen in Fig. \ref{fig:MRAsub1-8}, with numerical values in Table  \ref{Table3peerproceedings}.


The  "queen effect"   is well seen on the NJP data and fits, on Fig. \ref{fig:MRAmag}, but not so much on the CDF.
    The   "king effect"   is well seen on the NJP data and fits, on Fig. \ref{fig:MRAsup}, but the CDF shows a pronounced cut-off at high $r$. Therefore it would seem that the CDF is less pertinent to observe minute effects. This is understandable since the CDF results from an integration scheme. However, again understandably, the CDF fits are much more stable.

\bigskip
 
{\bf References}\label{biblio}
\bigskip
\vskip0.2cm 
Amati, G. \& van Rijsbergen, C. J.  (2002).  Term frequency normalization via Pareto distributions. {\it Advances in Information Retrieval},   F. Crestani, M. Girolami, C.J. van Rijsbergen (Eds.),  LNCS 2294, pp. 183--192. \vskip0.2cm 

Ausloos, M. (2013). A scientometrics law about co-authors and their ranking. The co-author. {\sl Scientometrics, 95}(3), 895-909.\vskip0.2cm 

Ausloos, M.  (2014). Binary Scientific Star Coauthors Core Size. {\it Scientometrics}, in press \vskip0.2cm

 Bar-Ilan, J. (2008). Which h-index? - A comparison of WoS, Scopus and Google Scholar. {\it Scientometrics,  74}(2),  257-271. \vskip0.2cm

    Benguigui, L. \&  Blumenfeld-Lieberthal, E. (2011). The end of a paradigm is Zipf's law universal?.   {\it  Journal of Geographical Systems, 13}(2), 87-100.     
 \vskip0.2cm 
  
Bonitz, M.,  \& Scharnhorst, A., (2011).    Competition in science and the Matthew core journals. {\it Scientometrics, 51}(1), 37--54. \vskip0.2cm 

Bonitz, M., Bruckner, E.,  \& Scharnhorst, A., (1999). The Matthew index-concentration patterns and Matthew core journals. {\it Scientometrics, 44}(3), 361-378.  \vskip0.2cm
 
 Bornmann, L., Mutz, R.,   \& Daniel, H. (2008). Are there better indices for evaluation purposes than the h-index? A comparison of nine different variants of the h-index using data from biomedicine. {\it Journal of the American Society for Information Science and Technology, 59}(5), 830-837.  \vskip0.2cm

 Bougrine,  H. (2014). Subfield Effects on the Core of Coauthors.   {\it  Scientometrics, 98}(2), 1047-1064. \vskip0.2cm

  Cole, J.R. \& Cole, S. (1972).  The Ortega Hypothesis Citation analysis suggest that only a few scientists contribute to scientific progress.  {\it  Science, 178}(4059), 368--375.  \vskip0.2cm
  
  Egghe,  L. (2005). {\it Power Laws in the Information Production Process Lotkaian Informetrics}. Amsterdam: Elsevier \vskip0.2cm

  Egghe, L.,   \& Rousseau, R. (1990). {\it Introduction to Informetrics. Quantitative Methods in Library, Documentation and Information Science}.  Amsterdam:  Elsevier \vskip0.2cm
 
   Fairthorne, R.A. (1969).   Empirical hyperbolic distributions (Bradford-Zipf-Mandelbrot) for bibliometric description and prediction. {\it Journal of Documentation, 25}(4),    319-343.  \vskip0.2cm

  Garfield, E. (2006). Citation indexes for science. A new dimension in documentation through association of ideas. {\it International journal of epidemiology, 35}(5),  1123--1127.   \vskip0.2cm 

  Glaeser, E.L. (2008). {\it Cities, agglomeration and spatial equilibrium}.  New York: Oxford University Press  \vskip0.2cm 
 
Haitun, S. D. (1982). Stationary scientometric distributions. Part. 1. Different approximations. {\it Scientometrics, 4}(1), 5-25. \vskip0.2cm

 Hirsch, J. E.   (2005). An index to quantify an individual's 
scientific research output. {\it Proceedings of the National Academy of Sciences USA,  102}(46), 16569-16572.  \vskip0.2cm 

 Hirsch, J. E.    (2010). An index to quantify an individual's scientific research output that takes into account the effect of multiple co-authorship. {\sl Scientometrics,  85}(3),   741-754.   \vskip0.2cm 
   
   Hsu, J. W. \& Huang, D. W. (2009). Distribution for the number of co-authors. {\it Physical Review E, 80}(5), 057101. \vskip0.2cm 
  
 Iszak, J. (2006). Some practical aspects of fitting and testing the Zipf-Mandelbrot model. {\it Scientometrics,    67}(1),  107--120.  \vskip0.2cm 
  
  Jarque, C.  M., \& Bera, A. K. (1980).  Efficient tests for normality, homoscedasticity and serial independence of regression residuals. {\it Economics Letters 6}(3), 255-259.   \vskip0.2cm  
  
  Jefferson, M. (1939)  The law of primate city. {\it Geographical Review, 29}(2), 226-232.  \vskip0.2cm 
  
        Laherr\`ere, J.  \&   Sornette,   D. (1998).   Stretched exponential distributions in nature and economy Òfat tailsÓ with characteristic scales. {\it European Physics Journal  B, 2}(4),     525-539.  \vskip0.2cm 
       
Madden C.H. (1958). Some Temporal Aspects of the Growth of Cities in the United States. {\it Economic Development and Cultural Change, 6}(2), 143-170.   \vskip0.2cm 

 Mandelbrot,  B. (1960). The Pareto-Levy Law and the Distribution of Income. {\it International Economics Review, 1}(2), 79-106. \vskip0.2cm

   Manin,   D. Yu. (2009). Mandelbrot's Model for Zipf's Law Can Mandelbrot's Model Explain Zipf's Law for Language?. {\it Journal of Quantitative Linguistics,   16}(3), 274--285. \vskip0.2cm
 
 Miskiewicz,  J. (2013). Effects of Publications in Proceedings  on the Measure of the Core Size of Coauthors.   {\it Physica A, 392}(20), 5119-5131. \vskip0.2cm
 
Newman, M.E.J.   (2001).  The structure of scientific collaboration networks. 
 {\it Proceedings of the National Academy of Sciences USA, 98}(2),    404-409.  \vskip0.2cm
 
Pareto, V. (1896) {\it Cours d'Economie Politique}. Geneva, Switzerland: Droz. \vskip0.2cm

Popescu, I. I., Altmann, G.,  K\"ohler, R. (2010). Zip's law - another view. {\it Quality and Quantity, 44}(4), 713-731.   \vskip0.2cm 

Price, D. J. de S.  (1956). The exponential curve of science.  {\sl Discovery, 17}(6), 240-243.   \vskip0.2cm 

Rosen, K.T \& Resnick, M. (1980). The size distribution of cities an examination of the Pareto law and primacy.  {\it Journal of Urban Economics, 8}(2), 165--186.    \vskip0.2cm 

 Rousseau, R. (2006). New developments related to the Hirsch index.  {\sl Science Focus, 1}(1), 23-25.   \vskip0.2cm 

  Schreiber, M. (2010). Twenty Hirsch index variants and other indicators giving more or less preference to highly cited papers.  {\it Annalen der Physik (Berlin), 522}(8), 536-554.
   \vskip0.2cm 

 Tsallis, C., \& Albuquerque, M.P. (2000). Are citations of scientific papers a case of nonextensivity?. {\it European Physics Journal  B, 13}(4), 777-780.   \vskip0.2cm 

   van Raan, A. F. J. (1996), Advanced bibliometric methods as quantitative core of peer review based evaluation and foresight exercises. {\it Scientometrics, 36}(3), 397-420.   \vskip0.2cm 
 
 Voloshynovska, I. A. (2011) Characteristic Features of Rank-Probability Word Distribution in Scientific and Belletristic Literature. {\it Journal of Quantitative Linguistics, 18}(3),  274-289. \vskip0.2cm 
 
 West, B.J. \& Deering, B. (1995)  {\it The Lure of Modern Science Fractal Thinking}.  Singapore: World Scient.    \vskip0.2cm 

 Zipf,   G.K.  (1949).  {\it Human Behavior and the Principle of Least Effort  An Introduction to Human Ecology. } Cambridge:  Addison Wesley.     \vskip0.2cm

\end{document}